\documentclass[aps,prd,superscriptaddress,nofootinbib,amsmath,amsfonts,groupedaddress]{revtex4-2}
\usepackage{graphicx}
\usepackage{setspace}
\usepackage{subfig}
\usepackage{epsfig}
\usepackage{bm}
\usepackage{amssymb}
\usepackage{float}
\usepackage{amsmath}
\usepackage{dcolumn}
\usepackage[colorlinks]{hyperref}
\usepackage[usenames,dvipsnames]{color}
\usepackage{enumitem}

\hypersetup{ breaklinks=true, pdfstartview={FitH}, colorlinks=true, linkcolor=blue, citecolor=red, filecolor=magenta, urlcolor=blue, anchorcolor=green, linktocpage=true }

\begin{document}

\title{ Power law cosmology in modified theory with thermodynamics analysis}

\author{J. K. Singh}
\email{jksingh@nsut.ac.in}
\affiliation{Department of Mathematics, Netaji Subhas University of Technology, New Delhi-110078, India}
\author{Shaily}
\email{shailytyagi.iitkgp@gmail.com}
\affiliation{Department of Mathematics, Netaji Subhas University of Technology, New Delhi-110078, India}
\affiliation{School of Computer Science Engineering and Technology, Bennett University, Greater Noida, India}
\author{Anirudh Pradhan}
\email{pradhan.anirudh@gmail.com}
\affiliation{Centre for Cosmology, Astrophysics and Space Science (CCASS), GLA University, Mathura-281 406, Uttar Pradesh, India}
\author{Aroonkumar Beesham}
\email{beeshama@unizulu.ac.za}
\affiliation{Department of Mathematical Sciences, University of Zululand,\\ Private Bag X1001, Kwa-Dlangezwa 3886, South Africa}
\affiliation{Faculty of Natural Sciences, Mangosuthu University of Technology, P O Box 12363, Jacobs 4026, South Africa}

\begin{abstract}
\begin{singlespace}

In this paper, we consider a cosmological model in $ f(R, G) $ gravity in a flat space-time, where $ R $ is the Ricci scalar and $ G $ is the Gauss-Bonnet invariant. The function $ f(R, G) $  is taken as a linear combination of $ R $ and an exponential function of $ G $. We analyze the observational constraints under a power law cosmology which depends on two physical parameters: the Hubble constant $ H_0 $ and the deceleration parameter $ q $. We constrain these two dependent parameters using the latest 77 points of the OHD data, 1048 points of the Pantheon data, and the joint data OHD+Pantheon and compare the results with the $ \Lambda $CDM. Also, we speculate constraints using a simulated data set for the future JDEM (Joint Dark Energy Mission)/Omega, supernovae survey. We see that $ H_0 $ is in very close agreement with some of the latest results from the Planck Collaboration that assume the $ \Lambda $CDM model. Our work in power law cosmology better fits the Pantheon data than the earlier analysis  \cite{Kumar:2011sw, Rani:2014sia}. However, the constraints obtained on $ H $ average, $ <H_0> $ and $ q $ average, $ <q> $ using the simulated data set for the future JDEM/Omega, supernovae survey are found to be inconsistent with the values obtained from the OHD and the Pantheon data. Additionally, we discuss statefinder diagnostics and see that the power law models approach the standard $\Lambda $CDM model ($ q\rightarrow -1 $). This model satisfies the Generalized Second Law of Thermodynamics. Finally, we conclude that the power law cosmology in $ f(R, G) $ gravity explains most of the distinguished attributes of evolution in cosmology.

\end{singlespace}
\end{abstract}

\maketitle
%PACS numbers: {98.80.-k, 95.36.+x, 95.35.+d, 04.50.-h}\\


\section{Introduction}

\qquad Current standard observations including type Ia Supernovae (SNIa), the cosmic microwave background (CMB) radiation, large scale structure (LSS), the Planck satellite, baryon acoustic oscillations (BAO), and the Wilkinson microwave anisotropy probe (WMAP) provide strong evidence about the accelerated expansion of the universe. It is noticed that modified gravity may describe the accelerated expansion of the universe in a better way. As we know, the modified gravity model is a simple gravitational alternative to the dark energy model. The idea behind these approaches to dark energy consists of adding additional gravitational terms to the Einstein-Hilbert action, which changes the universe's evolution in early or late times. Many examples of such models in modified gravity abound in the literature \cite{Cognola:2007vq, Nojiri:2007bt,  Bamba:2015jqa}. During the inflationary era, the Universe expanded at an extremely rapid rate. The inflationary era came to light during the late $70's$ in the early $ 80's $, which solved some of the problems of the Big Bang model. Bouncing cosmological models may be an acceptable description of the universe at early and late times and fit observations. This can be described by modified gravity in a unified way. To explain the accelerated expansion in standard general relativity, a phantom fluid or field is required. This phantom field eventually leads to a big rip, i.e., to a crushing type of singularity. \cite{Odintsov:2018nch}.

The late-time acceleration of the universe can also be described by modified gravity. If we replace the scalar curvature $ R $ in the Einstein-Hilbert action by $f(R)$, where $ f(R) $ is arbitrary, then we get $ f(R) $   gravity. This theory is simple, viable, and quite successful. There are many modifications of general relativity. If the Lagrangian is a function of both $ R $ and the trace $ T $ of the energy momentum tensor, then we get $ f(R,T) $ theory \cite{Alvarenga:2013syu, Sharif:2012zzd, Houndjo:2011tu, Jamil:2011ptc, Yousaf:2016lls, Godani:2018sbl, Alves:2016iks, Sharma:2014zya, Nagpal:2018uza, Yousaf:2017hsq, Das:2016mxq, Singh:2014kca, Nagpal:2018mpv}.  The $T$ term is introduced to take into account quantum effects, viscosity, and heat conduction. The late-time cosmic acceleration can also be explained in this gravity theory. $ f(R, T) $ gravity has been subjected to observational constraints. On the other hand, an interesting alternative to $ f(R) $ gravity is $ f(R, G) $. Here $G$ is the Gauss-Bonnet invariant constructed from the invariants $R_{\mu\nu}R^{\mu\nu} $ and $ R_{\mu\nu\alpha\zeta}R^{\mu\nu\alpha\zeta} $, where $R^{\mu\nu} $ is the Ricci tensor, and $R_{\mu\nu\alpha\zeta}$ is the Riemann tensor. In the literature, it is found that $ f(R, G) $ gravity is capable of describing inflation and late-time acceleration  \cite{ Li:2007jm, Nojiri:2005jg, Nojiri:2005am, Cognola:2006eg, Elizalde:2010jx, Izumi:2014loa, Oikonomou:2016rrv, Kleidis:2017ftt, Shaily:2024rjq, Oikonomou:2015qha, Escofet:2015gpa, Makarenko:2017vuk, Bamba:2014mya, Makarenko:2016jsy, MontelongoGarcia:2010ip}. Here, our main interest is to analyze the physical parameters of the universe in $ f(R, G) $ gravity.

The standard cosmological model has been incredibly successful in explaining the large-scale structure and history of the universe. However, it still has a major mystery such as the cosmological constant problem. This means we do not understand why the universe seems to be constantly expanding at an accelerating rate. Nevertheless, in the literature, many different models explain the main fearures of the universe. Models based on a power-law of the scale factor are quite successful in solving the age, horizon, and flatness problems that occur in the standard model \cite{Lohiya:1998tg, Sethi:1999sq, Batra:2000kur, Gehlaut:2002mj, Gehlaut:2003xi, Dev:2008ey, Dev:2002sz, Zhu:2007tm}. Sethi \textit{et al.} discussed an open linear coasting cosmological model based on a power law model \cite{Sethi:2005au}. Shafer used observation data and studied a robust model using a power law \cite{Shafer:2015kda}. Some remarkable works on other types of modified theories have been carried out by several authors. Recently, many excellent works have been accomplished by several authors in the alternative theory of gravity \cite{deHaro:2023lbq, Singh:2018xjv, Singh:2022jue, Aviles:2014rma, delaCruz-Dombriz:2016bqh, Capozziello:2019cav, Shaily:2024nmy, Balhara:2023mgj, Singh:2024ckh, Shaily:2024xho, Goswami:2023knh, Shaily:2022enj, Pawar:2024juv, Shabani:2016dhj, Singh:2022nfm}. Overall, power-law cosmology offers a promising alternative to the standard model, potentially addressing some of its shortcomings. However, further research is needed to explore its full implications.

The present work is organized as follows: In Sec.II, we evaluate the Einstein field equations for $f(R, G)$ gravity and using a power law cosmology, we have calculated the pressure and energy density in terms of the deceleration parameter $ q $. In Sec. III, we constrain the best-fit values of $ H_0 $ and $ q $ using MCMC simulation. Also, we predict constraints employing a simulated data set for the future JDEM, supernovae survey \cite{Holsclaw:2010sk, SNAP:2004hke} and perform Statefinder analysis of the obtained results. In Sec. IV, we discuss the cosmological parameters one by one and also observe the viability of the energy conditions. In the same section, the Statefinder and $ Om $ diagnostics are studied. In Sec. V, we measure the total entropy and Hawkings temperature for this model. And Finally, we summarize the outcomes of the obtained model.

\section{The Action and Cosmological Solutions}
\subsection{Field Equations}
\qquad The action of modified Gauss-Bonnet gravity in four dimension space-time is: \cite{MontelongoGarcia:2010ip, Odintsov:2018nch} 

\begin{equation} \label{1}
S=\int \left[ \frac{f(R,G)}{2\kappa}\right]  \sqrt{-g}d^{4}x+S_m,
\end{equation} 
where $ \kappa=8\pi G $, and $ S_m $ is the matter Lagrangian, which depends upon $g_{\mu\nu}$ and matter fields. The Gauss-Bonnet invariant $G$ is defined as $ G=R^2+R_{\mu\nu\alpha\zeta}R^{\mu\nu\alpha\zeta}-4R_{\mu\nu}R^{\mu\nu} $. The Gauss-Bonnet invariant is obtained from  $ R_{\mu\nu\alpha\zeta} $, $ R_{\mu\nu} = R^{\zeta}_{\mu\zeta\nu} $ and $ R = g^{\alpha\zeta}R_{\alpha\zeta} $.
From the equation of action (\ref{1}), the gravitational field equations  are derived as

\begin{multline}\label{2}
R_{\mu\nu}-\frac{1}{2} g_{\mu\nu}F(G)+(2RR_{\mu\nu}-4R_{\mu\alpha}R^\alpha_\nu+2R^{\alpha\zeta\tau}_\mu R_{\nu\alpha\zeta\tau}-4g^{i \alpha} g^{j \zeta}R_{\mu i \nu j}R_{\alpha\zeta})F'(G)
 +4[\nabla_\alpha \nabla_\nu F'(G)]R^\alpha_\mu \\ -4g_{\mu\nu}[\nabla_\alpha \nabla_\zeta F'(G)]R^{\alpha\zeta}+4[\nabla_\alpha \nabla_\zeta F'(G)]g^{i \alpha} g^{j \zeta}R_{\mu i \nu j}+2g_{\mu\nu}[\Box F'(G)]R
 -2[\nabla_\mu \nabla_\nu F'(G)]R \\ -4[\Box F'(G)]R_{\mu\nu}+4[\nabla_\mu \nabla_\alpha F'(G)]R^\alpha_\nu=\kappa T^m_{\mu\nu},
\end{multline}

where $ T^m_{ij} $ is the energy momentum tensor arising from $ S_m $. The flat FLRW space-time metric is:

\begin{equation} \label{3}
ds^{2}=-dt^{2}+a^{2}(t)(dx^{2}+dy^{2}+dz^{2}),  
\end{equation}
where the symbols have their usual meanings.  Now, we calculate the Einstein field equations using Eqs. (\ref{2}) and (\ref{3}) as:

\begin{equation}\label{4}
F(G)+6H^2-GF'(G)+24H^3 \dot{G} F''(G)=2\kappa \rho,
\end{equation}

\begin{multline}\label{5}
6H^2+4\dot{H} +F(G)+16 H\dot{G}(\dot{H}+H^2)F''(G)-G F'(G)++8H^2\ddot{G} F''(G)+8H^2\dot{G}^2F'''(G)=-2\kappa p,
\end{multline}
Here $ H=\frac{\dot{a}(t)}{a(t)} $ is the Hubble parameter and $\dot{a}(t) \equiv \frac{da}{dt}$.  Also, we have
\begin{equation} \label{6}
R=6(2H^2+\dot{H}),
\end{equation}
\begin{equation} \label{7}
G=24H^2(H^2+\dot{H}).
\end{equation}
In the present model, we are taking $ F(R, G)=R+ \alpha e^{-G} $ and this term denotes the difference with general relativity.  Here, $ \alpha $ is an arbitrary positive constant and if $\alpha=0$ then it will return to GR theory.

\subsection{Power law cosmology}

To implement power-law cosmology, we take the scale factor $ a(t) $ as \cite{Rani:2014sia, Kumar:2011sw}
\begin{equation} \label{8}
a(t)=a_0(\frac{t}{t_0})^{\zeta},
\end{equation}
where $ a_0 $ is the value of $ a(t) $ at present, and $ \zeta $ is a parameter which is dimensionless. Now, the Hubble parameter can be described using the scale factor as 
\begin{equation} \label{9}
H \equiv \frac{\dot{a}}{a}=\frac{\zeta}{t}. 
\end{equation}
Also
\begin{equation} \label{10}
H_0=\frac{\zeta}{t_0}.
\end{equation}

Now, since we know the relation between the scale factor and redshift \text{i.e.} $ a(t)=\frac{a_0}{1+z}$, where $z$ is the redshift, $H$ can be in terms of  $ (z) $ as

\begin{equation} \label{11}
H(z)=H_0(1+z)^{\frac{1}{\zeta}}.
\end{equation}

To understand the history of the universe, we consider cosmological parameters like pressure, energy density,  EoS parameter, Hubble parameter, deceleration parameter, etc. The acceleration or deceleration phase of the universe can be measured by a dimensionless quantity which is known as the deceleration parameter. The deceleration parameter $ q $  is defined as:

\begin{equation} \label{12}
	q=-\frac{\ddot{a}}{aH^2}.
\end{equation}  

Now if $ q>0 $, we have a decelerating universe, if $ q < 0 $, then it indicates acceleration phase and if $ q = 0 $, then the expansion will be at a constant rate. 
Eqs. (\ref{8}), (\ref{9}) and (\ref{12}) yield 

\begin{equation}\label{13}
q=\frac{1}{\zeta}-1.
\end{equation}

Thus, we represent the Hubble parameter in terms of deceleration parameter $ q $ and redshift as

\begin{equation} \label{14}
H(z)=H_0(1+z)^{(1+q)}.
\end{equation}

The energy density and the pressure can be obtained by solving Eqs. (\ref{4}) and (\ref{5}) which are given as

\begin{equation} \label{15}
\rho=\frac{\alpha e^{24 H_0^4 q (z+1)^{4 q+4}} (24 H_0^4 q (z+1)^{4 q+4}(96 H_0^4 (q+1) (z+1)^{4 q+4}-1)+1)+6 H_0^2 (z+1)^{2 q+2}}{2 \kappa },
\end{equation}
\begin{multline} \label{16}
p=\frac{1}{2 \kappa}\Bigg[-192 \alpha H_0^4 (z+1)^{4 q+4} e^{24 H_0^4 q (z+1)^{4 q+4}} (H_0 (z+1)^{q+1})^{3}+16 H_0^4 (q+1) (q (11 q+4)-3) (z+1)^{4 q+4}+q)-1)+ \\ \alpha e^{24 H_0^4 q (z+1)^{4 q+4}} (24 H_0^4 (z+1)^{4 q+4} (3072 H_0^8 q^2 (q+1)^2 (z+1)^{8 q+8} +2 H_0^2 (2 q-1) (z+1)^{2 q+2}\Bigg],
\end{multline}
\begin{equation} \label{17}
\omega=\frac{p}{\rho}.
\end{equation}

For further analysis, we take $\alpha$ and $\kappa$ equal to unity and constrain the model parameters $ H_0 $ and $ q $ using recent observational data sets.

\section{Observational constraints}
\qquad In this section, observational data sets are used to constrain the value of $ H_0 $ and $ q $ which appear in the tilted Hubble parametrization. In the present model, we use the $ H(z) $, Pantheon datasets, and their joint data set.

\subsection{OHD Data set}
Here, we use OHD (77 points) as compiled by Shaily  \cite{Shaily:2022enj}. Now, best-fit values of $ H_0 $ and $ q $ are obtained from the usual chi-square test.  Chi-square is given by:

\begin{equation}\label{23}
\chi _{HD}^{2}(H_0,q)=\sum\limits_{i=1}^{77} \frac{[H(H_0,q,z_{i})-H_{obs}(z_{i})]^2}{\sigma _{z_i}^2},
\end{equation}

where $ H_{obs}$ and $ H(H_0,q,z_{i}) $ are the observed and theoretical values, respectively and $ \sigma_{(z_{i})} $ is  the standard deviation at $ H(z_i) $.  

\begin{figure}
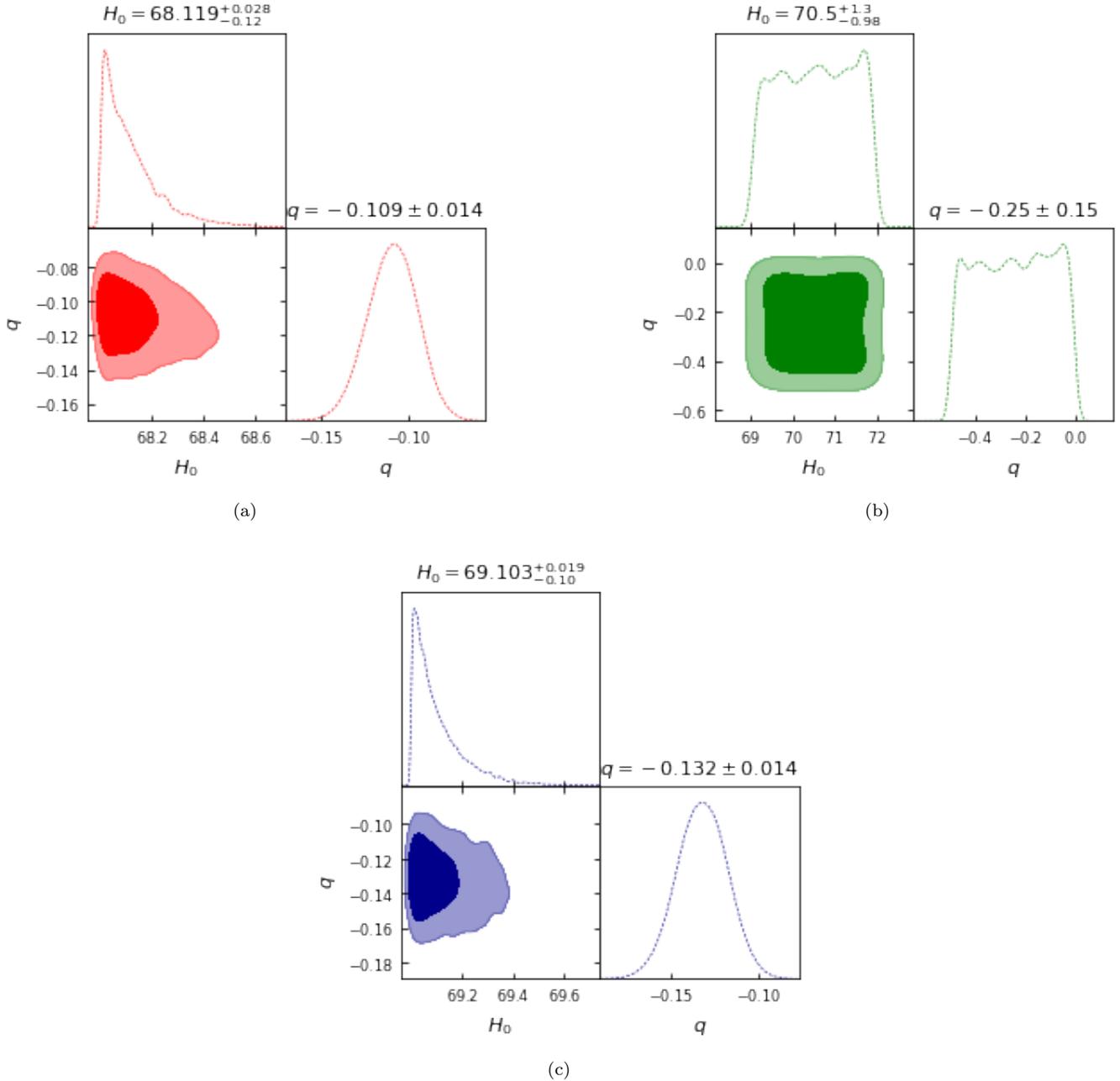
\centering
	\subfloat[]{\label{hpca}\includegraphics[scale=0.7]{HC}}\hfill	
	\subfloat[]{\label{hpcb}\includegraphics[scale=0.7]{PC}}\par	
	\subfloat[]{\label{hpcc}\includegraphics[scale=0.7]{HPC}}	
\caption{\scriptsize Figs. show the $ H_0 $-$ q $ likelihood contours for $ H(z) $, Pantheon and $ H(z)+Pantheon $ data set.}
\label{hpc}
\end{figure}

\subsection{Pantheon Data set}

\qquad We use Pantheon compilation, which consists of $1048$ data points in the redshift range $ 0.01 < z < 2.26 $. This data is collected from different supernovae survey \textit{e.g.}  CfA 1-4, CSP, SDSS, SNLS, PS1, and high-z in the redshift range $ 0.01 \leq z \leq 2.26 $ according as the Table \ref{tabpan1}. The SNIa data plays a key role in investigating the expansion rate of the universe \cite{Asvesta:2022fts}. 

\begin{table}[htbp]
\caption{ {\bf Pantheon Compilation}}
%\begin{ruledtabular}
\begin{center}
\label{tabpan1}
\begin{tabular}{l c c c} 
\hline\hline
\\ 
{Survey} &      ~~~~~  Data Points  & ~~~~~ Redshift range &   ~~~~ Reference
\\
\\
\hline      
\\
{CfA 1-4 }     & ~~~~~ $ 147 $   &  ~~~~~ $ 0.01-0.07 $  &  ~~~~~~~~ \cite{Riess:1998dv, Jha:2005jg, Hicken:2009df, Hicken:2009dk, Hicken:2012zr}
\\
\\
{CSP }     &  ~~~~~ $ 25 $   &  ~~~~~ $ 0.01-0.06 $  &  ~~~~~~~~ \cite{Contreras:2009nt}
\\
\\
{SDSS }  & ~~~~~ $ 335 $   &  ~~~~~ $ 0.03-0.40 $  &  ~~~~~~~~ \cite{SDSS:2014irn}
\\
\\
{SNLS }  & ~~~~~ $ 236 $   &  ~~~~~ $ 0.12-1.06 $  &  ~~~~~~~~ \cite{SNLS:2010pgl}
\\
\\
{PS1 }  & ~~~~~ $ 279 $   &  ~~~~~ $ 0.02-0.63 $  &  ~~~~~~~~ \cite{Pan-STARRS1:2017jku}
\\
\\
{high-$z$ }  & ~~~~~ $ 26 $   &  ~~~~~ $ 0.73-2.26 $  &  ~~~~~~~~ \cite{SupernovaCosmologyProject:2011ycw, Riess:2017lxs, SupernovaSearchTeam:2004lze, Riess:2006fw}
\\
\\
\hline
\\
{Total }  & ~~~~~ $ 1048 $   &  ~~~~~ $ 0.01-2.26 $
\\
\\
\hline\hline  
\end{tabular}    
\end{center}
\end{table}

We compute the value of distance modulus $ \mu_{th}(z_i) $ using the predicted apparent magnitude (m) and absolute magnitude (M) \textit{w.r.t.} the color and stretch as
\begin{equation}\label{24}
\mu_{th}(z)= -M + m = \mu_{0} + 5\log D_L(z),
\end{equation}
where $ D_L(z) $ is the luminosity distance, and $ \mu_0 $ is the nuisance parameter. These  are given by:
\begin{equation}\label{25}
D_L(z)=c D_n(1+z)  \int_0^z \frac{1}{H(z^*)}dz^*,
\end{equation}
where
\begin{equation}\label{26}
D_n(z)=\begin{cases}
\frac{\sinh(\sqrt{\Omega_m})}{H_0 \sqrt{\Omega_m}}, \text{for} ~~\Omega_m>0\\
1, \text{for}~~~~~~~ \Omega_m=0\\
\frac{\sin(\sqrt{\Omega_m})}{H_0 \sqrt{\Omega_m}}, \text{for} ~~\Omega_m<0
\end{cases} 
\end{equation}
and
\begin{equation}\label{27}
\mu_0= 5 \log\Big(\frac{H_0^{-1}}{1Mpc}\Big)+25,
\end{equation}
respectively. \\

Now, the minimum $ \chi^2 $ function is given as

\begin{equation}\label{28}
\chi _{Pan}^{2}(H_0,q)=\sum\limits_{i=1}^{1048}\left[ \frac{\mu_{th}(H_0,q,z_{i})-\mu_{obs}(z_{i})}{\sigma _{\mu(z_{i})}}\right] ^2.
\end{equation}
where PAN stands for the observational Pantheon data set, $ \sigma_\mu (z_i) $ indicates the observed value's standard error, $ \mu_{th} $ the theoretical distance modulus, and $ \mu_{obs} $  the model's  observed distance modulus. 

\subsection{Joint Data set (OHD+Pantheon)}
\qquad By performing joint statistical analysis using $ H(z) $ and Pantheon data sets, we can obtain stronger constraints. Therefore, the chi-square function for joint data sets can be written as 
\begin{eqnarray}\label{29}
\chi_{Joint}^2= \chi_{HD}^2 + \chi_{PAN}^2.
\end{eqnarray}

\begin{table}
	\caption{Best-fit values of the Physical parameters }
	%\begin{ruledtabular}
	\begin{center}
		\label{tab1}
		\begin{tabular}{l c c c r} 
			\hline\hline
			\\
			{Physical parameters} &  ~~~~~~~  OHD & ~~~~~   Pantheon  & ~~~~~~    OHD+Pantheon\footnote{ transition occurs from deceleration to acceleration stated in Fig. \ref{sfb}.}\footnote{ The model represents the quintessence dark energy model at present.}\footnote{ the present value of the deceleration parameter is $ q_0\approx -0.54 $ with the recent observations \cite{Mamon:2016dlv}. }  & ~~~~~~    $ \Lambda $CDM$^*$\footnote{according to the Planck $ 2018 $ results. VI.  \cite{Planck:2018vyg}}
			\\
			\\
			\hline
			\\      
			{ $ H_0 $ }  &   ~~~~~~~ $ 68.119_{-0.12}^{+0.028} $   &  ~~~~~ $ 70.5_{-0.98}^{+1.3} $ & ~~~~~ $ 69.103_{-0.10}^{+0.019} $ &  $ 72.150_{+0.989}^{-0.779} $
			\\
			\\
			{ $ q $ }    & ~~~~~~~ $ -0.109_{-0.014}^{+0.014} $   &  ~~~~~ $ -0.25_{-0.15}^{+0.15} $ &  ~~~~~ $ -0.132_{-0.014}^{+0.014} $  & $ -0.533_{+0.024}^{-0.024} $  
			\\
			\\
			{ $\omega $ }  &  ~~~~~~~$ -0.406  $  & ~~~~~ $  -0.5 $ &  ~~~~~ $ -0.421 $ & $ -0.689_{+0.016}^{-0.016} $ 
			\\
			\\ 
   { $ s $ }    & ~~~~~~~ $ 0.594_{-0.009}^{+0.009} $   &  ~~~~~ $ 0.5_{-0.10}^{+0.10} $ &  ~~~~~ $ 0.579_{-0.009}^{+0.009} $ &
   $ 0.311_{+0.016}^{-0.016} $			\\
			\\
   { $ r $ }    & ~~~~~~~ $ -0.085_{-0.008}^{+0.008} $   &  ~~~~~ $ -0.125_{-0}^{+0} $ &  ~~~~~ $ -0.097_{-0.007}^{+0.007} $  & 
   $ 0.035_{+0.027}^{-0.027} $			\\
			\\
			\hline\hline  
		\end{tabular}    
		%\end{ruledtabular} 
	\end{center}
\end{table}
\begin{figure}\centering
	\subfloat[]{\label{era}\includegraphics[scale=0.43]{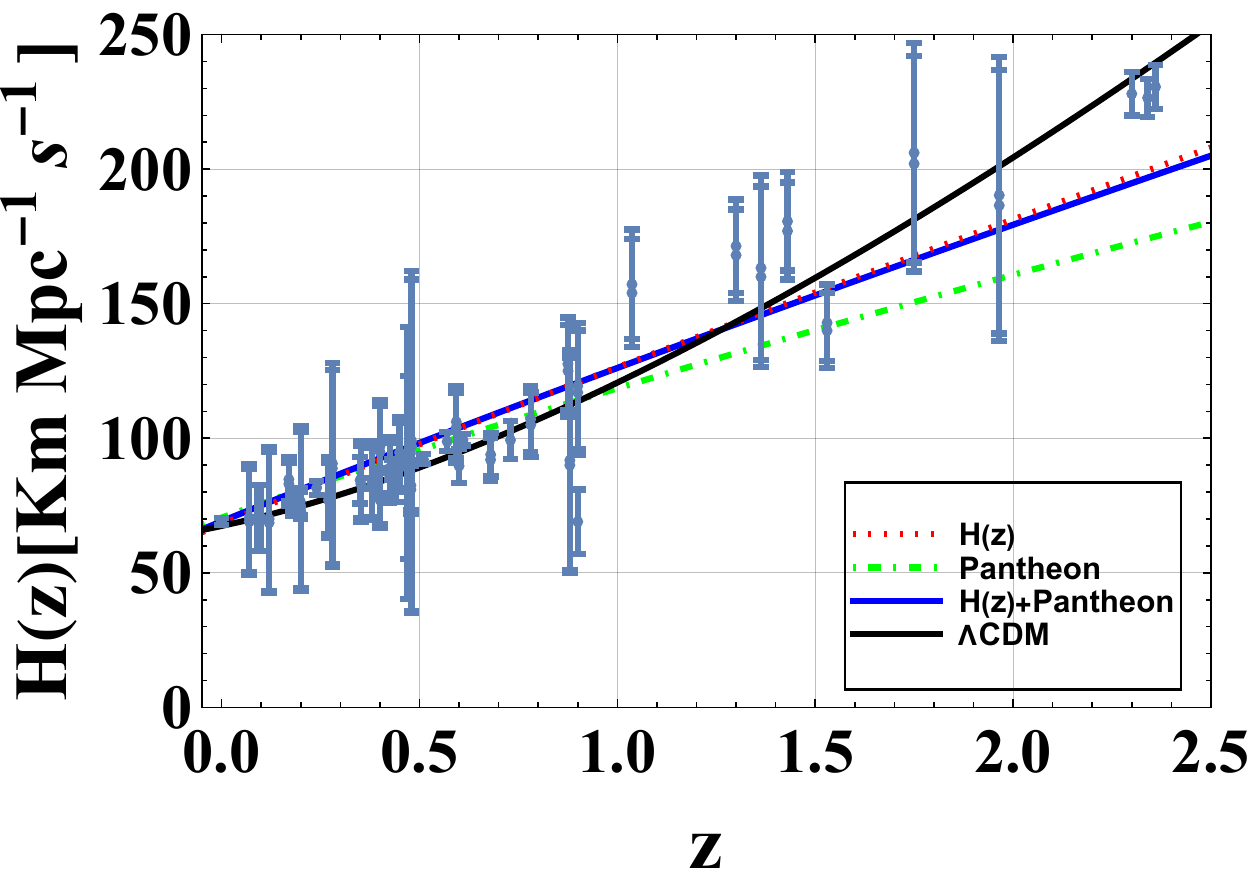}}\hfill
	\subfloat[]{\label{erb}\includegraphics[scale=0.40]{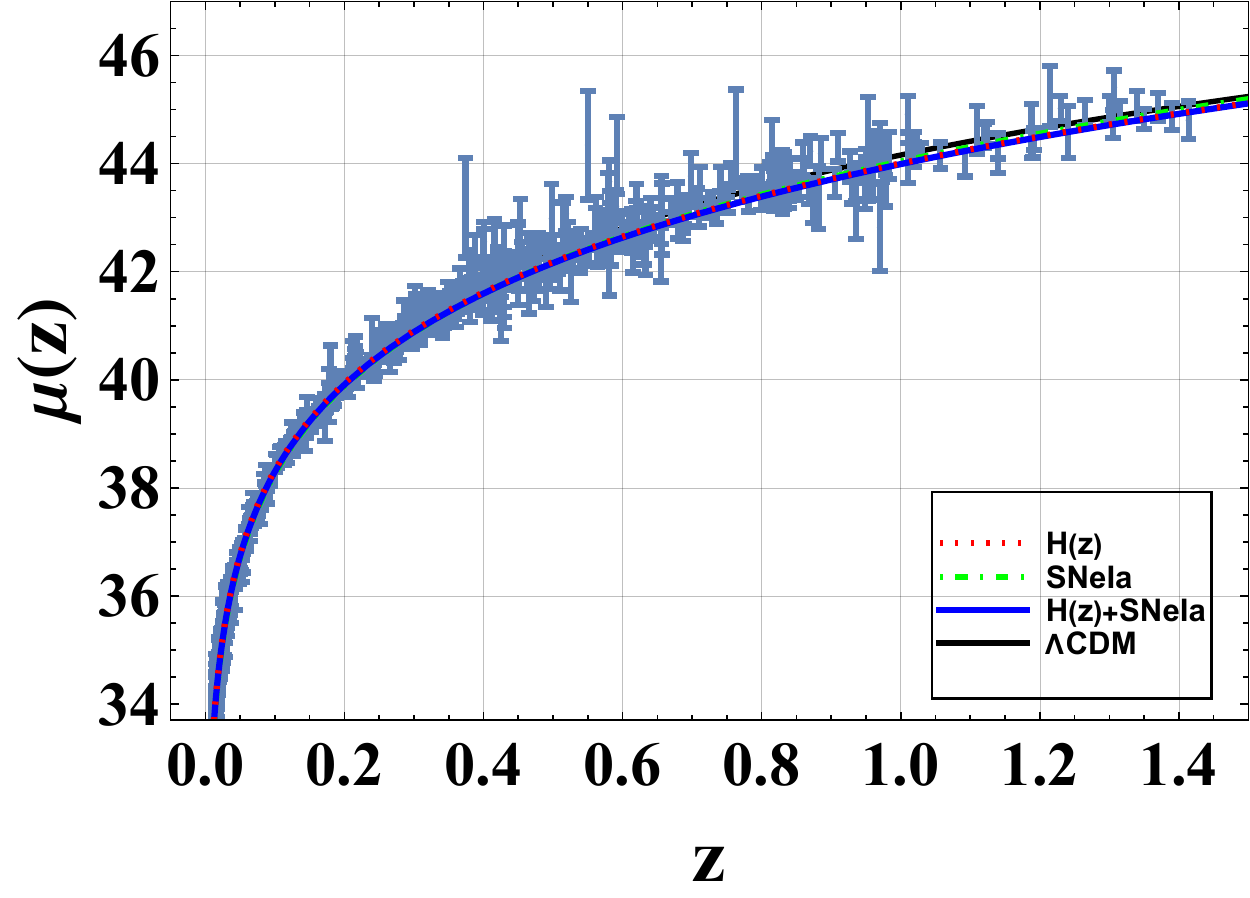}} 
\caption{ The error bar plots for OHD and SNIa data sets show the resemblance between our model and $ \Lambda $CDM.}
\label{er}
\end{figure}

%%%%%%%%%%%
\subsection{ $ \Lambda $CDM}

In standard cosmology, the cosmological constant $ \Lambda $ known as $ \Lambda $CDM is the simplest dark energy candidate whose energy density remains constant {\it w.r.t.} time given by $ \rho_\Lambda \equiv \frac{\Lambda}{8\pi G} = -p_\Lambda $, and $\omega_\Lambda=-1$ is its equation of state (EoS). The Hubble parameter for $ \Lambda $CDM model for flat FLRW Universe is given by
\begin{equation}\label{27a}
H(z)=H_0[\Omega_{m0} (1+z)^3+(1-\Omega_{m0})]^{1/2}.
\end{equation}
Here, $ \Omega_{m0} $ and $ H_0 $ are the present matter density and Hubble parameter respectively. 

\subsection{ JDEM: Simulated SNIa data }

In addition, we use a simulated dataset based on the upcoming JDEM SN-Survey with around $ 2300 $ SNIa data points in the redshift range from $ 0\leq z \leq 1.7 $ to analyze our future accomplishment \cite{Holsclaw:2010sk}.  We consider a simplified error model with identical errors for all supernovae independent of redshift. In this model, We assume a statistical error of $ \sigma = 0.13 $ magnitude, as expected from JDEM-like future surveys \cite{SNAP:2004hke}. A set of $ 500 $ data points has been generated using the latest specifications \cite{Alam:2003sc}, taking $ \Lambda $CDM model as our standard model with $ \Omega_{m0} = 0.3 $. The best-fit parameters $ H_0 $, $ q $, $ r $, and $ s $ were calculated using $ 500 $ data points. Finally, we computed the mean values of $ H_0 $, $ q $, $ r $, and $ s $ denoted as $ <H_0> $, $ <q> $, $ <r> $, and $ <s> $ respectively \cite{Rani:2014sia}. The numerical results are given in Table \ref{tabjdem}.

\section{ Physical consequences of the model}

\qquad For a flat universe, we evaluate the best-fit values of $ H_0 $ and $ q $ for the $ H(z) $, Pantheon, and their joint data sets, respectively. For this purpose, we perform the coding in Python, where we use the Monte Carlo Markov chain  (MCMC) method which is given in the  Python module \textit{emcee}, and plot the 2-D plots with $ 1-\sigma $, $ 2-\sigma $ likelihood contours. In Fig. \ref{hpc}, the contours are closed but are not in proper oval shape. The convergence of this type of plot also exists and can be measured using the Gelman-Rubinn convergence test. This test was used on a large scale in Bayesian inference to examine the convergence of the profile \cite{Singh:2022nfm, Gelman:1992zz, Brooks:1998}. For the  $ H(z) $ data set, the best fit values are $ H_0=68.119_{-0.12}^{+0.028} $ and $ q=-0.109_{-0.014}^{+0.014} $ (see Fig. \ref{hpca}). In Fig. \ref{hpcb}, we can see that for the Pantheon data set, the best fit values are $ H_0=70.5_{-0.98}^{+1.3} $ and $ q=-0.25_{-0.15}^{+0.15} $. For the joint data set, we obtain the best-fit value of $ H_0=69.103_{-0.10}^{+0.019} $ and $ q=-0.132_{-0.014}^{+0.014} $, which is observed in Fig. \ref{hpcc}. In our work, we notice that  $ q_0 $ differs from is not very close to -0.5, which is the approximate value for the  $\Lambda$CDM model. In the refs \cite{Singh:2018cip, Sahu:2016ccd}, it is pointed out that modified gravity theories could admit different values for $ q_0 $.  We plot the error bar plots for the Hubble and SNIa data sets with these best-fit values. In Fig. \ref{er}, one can compare the present model with the $ \Lambda $CDM model. 

\begin{table}
%\textbf{Table 2.} Summary of the numerical results.\\
\caption{ Simulated data of JDEM Survey.}
\begin{center}
\label{tabjdem}
\begin{tabular}{l c c c c c r }
\\
\hline\hline
Data & $<q>$ & $<H_{0}>$ \footnote{ $H$ is taken in the units of km/s/Mpc} & $<\omega_0>$ & $<r>$ & $<s>$
\\
\hline
\\
JDEM & $-0.0656_{+0.0003}^{-0.0003}$ & $25.9083_{+0.0030}^{-0.0030}$ &-& $-0.05699_{+0.00022}^{-0.00022}$ & $0.6229_{+0.0002}^{-0.0002}$
\\
\\
\hline\hline
\end{tabular}
\end{center}
\end{table}

\begin{figure}\centering
	\subfloat[]{\label{rhoa}\includegraphics[scale=0.5]{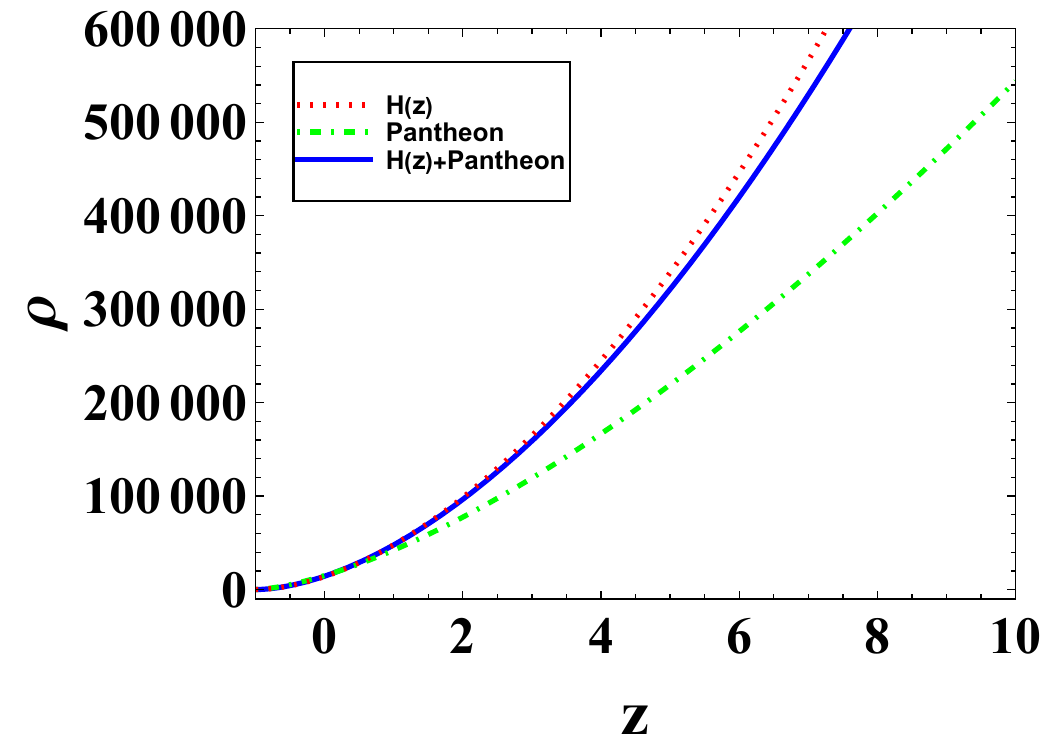}}\hfill
	\subfloat[]{\label{rhob}\includegraphics[scale=0.5]{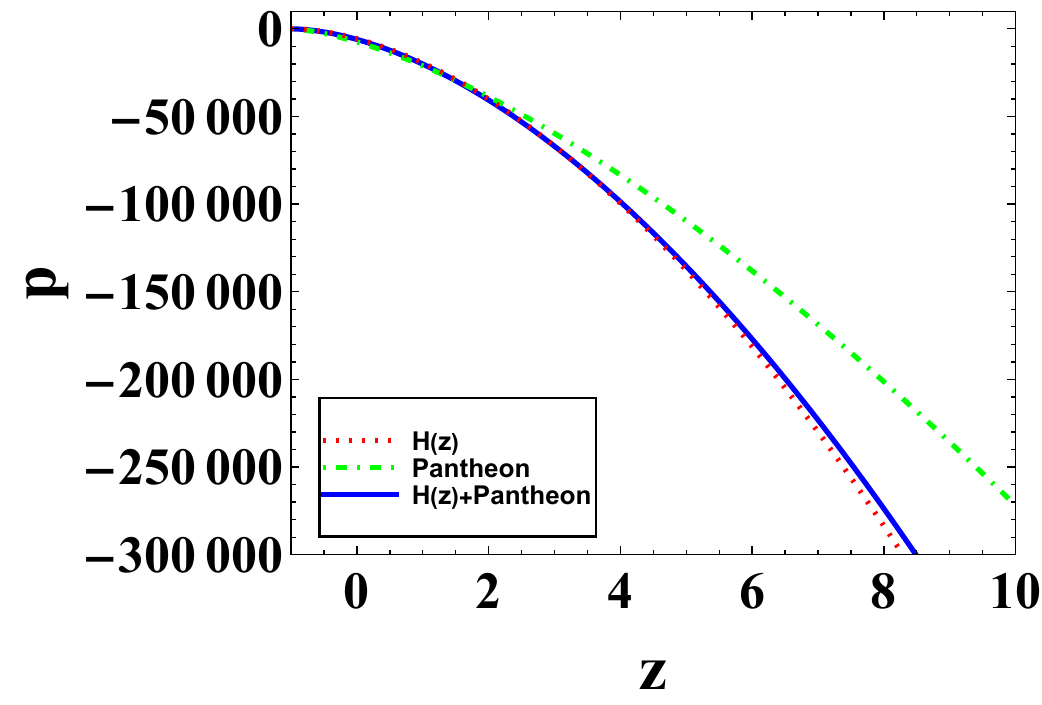}}
	%\subfloat[]{\label{rhoc}\includegraphics[scale=0.42]{EoS}}
\caption{ The plots of  $ \rho $,  $ p $ and $\omega$ against redshift $ z $}
\label{rho}
\end{figure}
\subsection{ Physical Parameters}

This section analyzes the changing behavior of energy density $ \rho $ and pressure $ p $. From Eqs. (\ref{15}) and (\ref{16}), it is clear that the values of $ \rho $ and $ p $ are in the terms of $ z $ and the value of $ \omega $ is in terms of $ q $ only, \textit{i.e.}, a constant. So, to understand the evolution of these parameters we plot the graphs. Fig. \ref{rhoa} shows the evolution of the energy density against the redshift. For high redshift $ \rho $ is very large, and as $ z $ decreases, $ \rho $ also decreases for the whole range of $ z $ ans as $ z \to -1 $, the energy density $ \rho \to 0 $. Further, Fig. \ref{rhob} shows the evolution of pressure $ p $ against redshift $ z $, and we observe that the pressure is negative, corresponding to accelerated expansion. From Eqs. (\ref{15}), (\ref{16}) and (\ref{17}), we calculate the value of the EoS parameter which is almost constant from high redshift to low redshift. The numerical values of $ \omega $ are measured as -0.406, ~-0.5, ~-0.421 for the $ H(z) $, Pantheon and their joint data sets, respectively. These EoS parameter values show that our model is in the quintessence region.
%%%%%%%%%%%%%%
\subsection{ Energy Conditions}
%%%%%%%%%%%%%%

\noindent The energy conditions (ECs) are relevant to study the issues related to the space-time singularity and the behavior of the null, space-like, time-like, or light-like geodesics. These conditions are discussed as a simple constraint on the linear combination of the energy density and the pressure. We study the energy conditions for the $ f(R, G) $ modified gravity \cite{MontelongoGarcia:2010ip}. This results in energy density not being negative and that gravity always indicates an attractive force. ECs always inflict restrictions on the capability of the stress tensor to contract at each location of the space \cite{Curiel:2014zba}. These energy conditions can be expressed in the geometric form in addition to the physical form which complies with the Ricci tensor as the stress tensor in the Einstein field equation for different theories of gravity. The ECs are explained in terms of the weak energy condition (WEC), the strong energy condition (SEC), the dominant energy condition (DEC), and the null energy condition (NEC), which are given in Table \ref{tab:2}, where $t^i$ and $\xi^j$ are co-oriented time-like vectors and $k^i$ is a null (light-like) vector. These conditions are dependent on each other \cite{Kontou:2020bta}. 

\begin{table}[htbp]
\caption{ \textbf{Energy Conditions}}
\centering
\begin{tabular}{l c c c c r}
\hline\hline
\\
Energy Condition & \qquad Physical form & \qquad Geometric form & \qquad Perfect Fluid form
\\
 \\
\hline
\\
 ~~ $ NEC $ ~~ & ~~ $ T_{ij}k^ik^j\geq 0 $ & ~~$ R_{ij}k^ik^j\geq 0 $  &  ~~ $ \rho+p \geq 0 $
\\
\\ 
~~ $ WEC $ ~~ & ~~ $ T_{ij}t^it^j\geq 0 $  & ~~$  G_{ij}t^it^j\geq 0 $  & ~~$  \rho \geq 0, \rho+p \geq 0 $  
 \\
\\
~~ $ SEC $ ~~ & ~~  $ (T_{ij}-\frac{T}{n-2}g_{ij})t^it^j\geq 0  $ & ~~ $ R_{ij}t^it^j\geq 0 $  & ~~  $ \rho+p \geq 0, (n-3)\rho+(n-1)p \geq 0 $
\\
\\
~~ $ DEC $~~  & ~~ $ T_{ij}t^i\xi^j\geq 0 $  & ~~ $ G_{ij}t^i\xi^j\geq 0 $ & ~~   $\rho \geq |p| $  
\\
 \\
\hline\hline
\end{tabular}
\label{tab:2}
\end{table}

In Table \ref{tab:2}, it is noted that if WEC, SEC, and DEC hold then NEC holds, but if the NEC does not satisfy then none of the ECs can be satisfied \cite{Visser:1995cc}. It is reviewed that the NEC is the viable condition for all of the stable models of the universe. In a sensible model, other ECs can be violated by adding a suitable (positive or negative) cosmological constant. If the stress-energy tensor violates the NEC, the system becomes catastrophically unstable, meaning the presence of the ghost system. This occurs when the system either with \textit{wrong} sign of the kinetic term or the system contains \textit{tachyons} \cite{Caldwell:1999ew}, which are in the modes that expand exponentially with arbitrarily short wavelengths. The modified gravity theories deviating from general relativity can be viewed as having a background of a scalar field with a stress-energy tensor and serving as a reasonably cosmological constant. In such theories, the NEC can be violated. As a result, the modified gravity theories may affect degenerating dispersion relations, enabling the breaking of the NEC.

\begin{figure}\centering
	\subfloat[]{\label{eca}\includegraphics[scale=0.47]{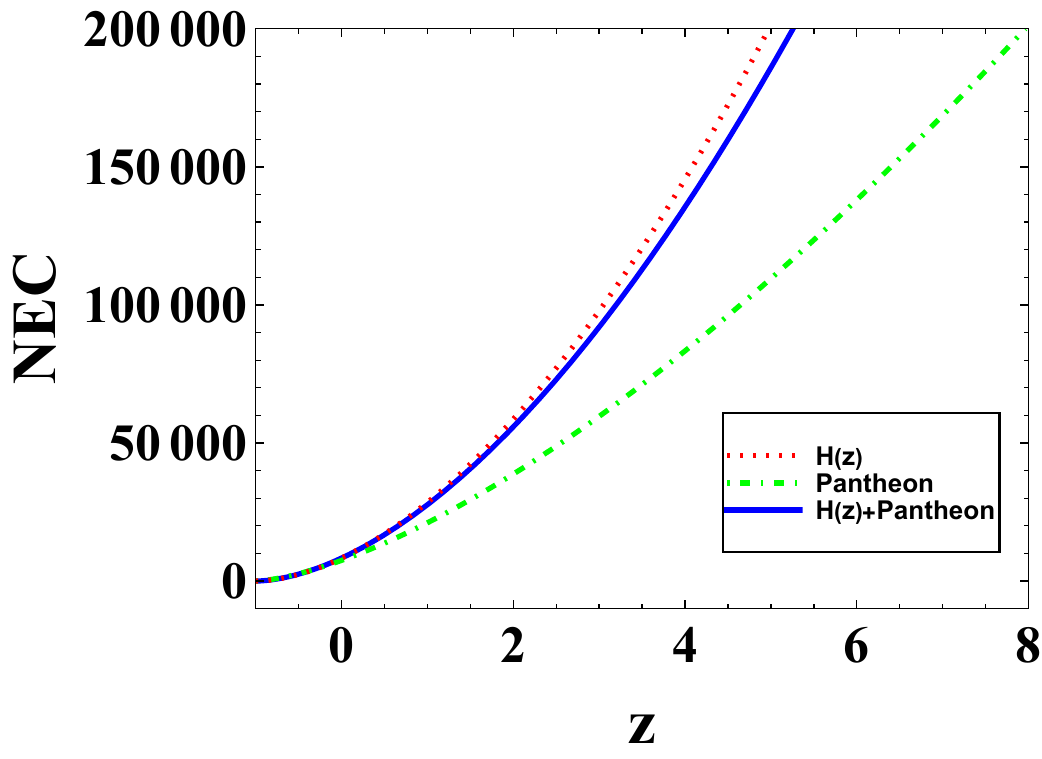}}\hfill
	\subfloat[]{\label{ecb}\includegraphics[scale=0.48]{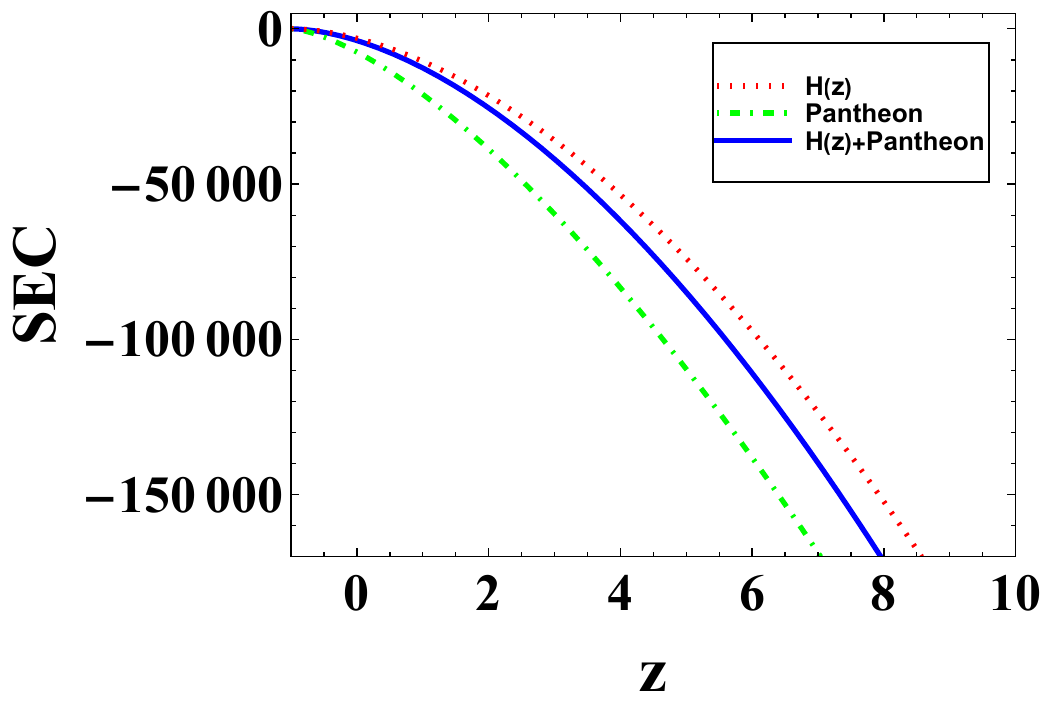}}\par
	\subfloat[]{\label{ecc}\includegraphics[scale=0.48]{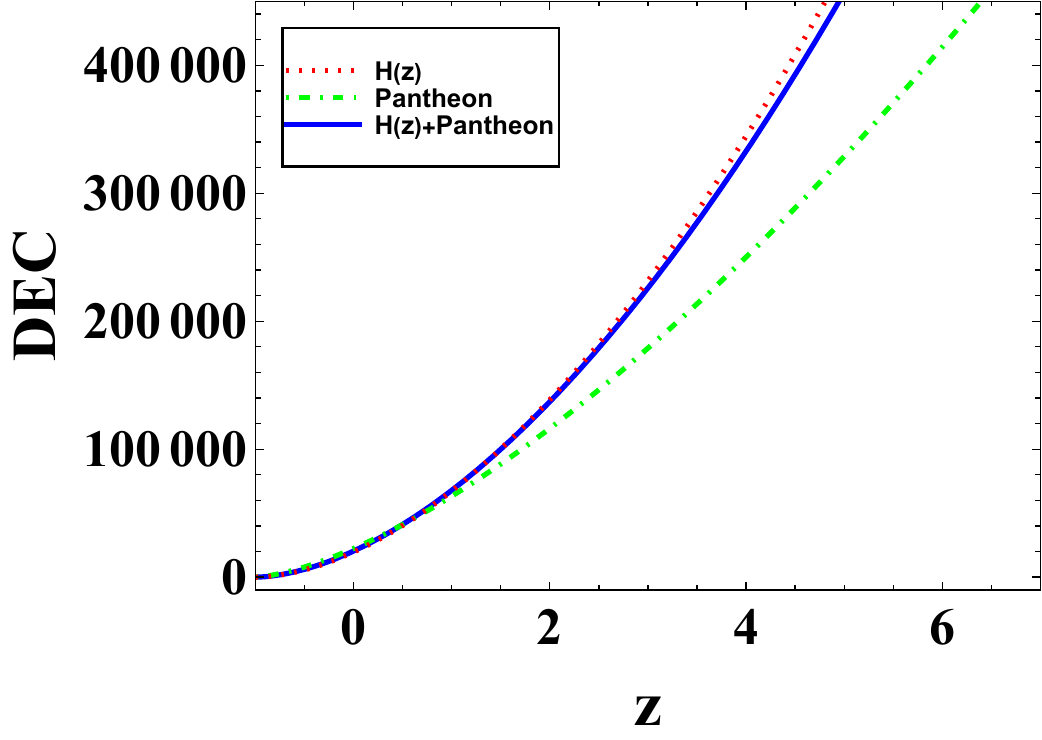}}
\caption{ The Plots of  NEC, WEC, DEC and SEC}
\label{ec}
\end{figure}
The NEC plays a prominent role among different ECs studied in general relativity. In the NEC, the energy-momentum tensor for matter $ T_{ij} $ satisfies $ T_{ij}k^i k^j $, $ \forall $ $ k^i $, \textit{i.e.}, for any vector with $ g_{ij}k^i k^j=0 $.
The main reasons why the NEC is contemplated significant \cite{Rubakov:2014jja}, (i) It was previously believed that the NEC could only be violated in a theory that involves a scalar field with non-minimal coupling to gravity \cite{Flanagan:1996gw}, (ii) the NEC is a fundamental assumption in the Penrose singularity theorem \cite{Penrose:1964wq}, which is applicable in general relativity. The Penrose singularity theorem is based on two conditions: (i) the NEC be upheld; (ii) the Cauchy hypersurface be non-compact. This theorem states that a trapped surface exists in space, and a singularity will eventually occur. The trapped surface is defined as a closed surface where the outgoing light rays converge (moving inward). 

In Fig. \ref{ec}, it is demonstrated that the NEC and DEC are satisfied, but the SEC is violated, indicating the universe's accelerated expansion. 

\subsection{Cosmographic parameters}

\qquad To understand the universe's expansion history, many cosmological parameters are studied, which are expressed in the form of higher-order derivatives of the scale factor. Therefore, to explore the dynamics of the universe, these parameters are very helpful. For example, Hubble parameter $ H $ shows the expansion rate of the universe, the deceleration parameter $ q $ tells about the phase transition of the universe, the jerk parameter $ j $, the snap parameter $ s $, and the lerk parameter $m$ investigate dark energy models and their dynamics. These are defined as:

\begin{equation} \label{18}
H=\frac{\dot{a}}{a};~~q=-\frac{\ddot{a}}{aH^{2}};~~j=\frac{\dddot{a}}{aH^{3}}%
;~~s=\frac{\ddddot{a}}{aH^{4}};~~l=\frac{\ddddot{\dot{a}}}{aH^{5}};~~m=\frac{\ddddot{\ddot{a}}}{aH^{4}}.
\end{equation}  

These parameters may also be written in terms of $q$ as

\begin{equation} \label{19}
j=q(1+2q); ~~s=-q(1+2q)(2+3q);~~l=q(2+3q)(1+2q)(3+4q); ~~m=-q(2+3q)(1+2q)(3+4q)(4+5q).
\end{equation}
Here $ j $, $ s $, $ l $, and $ m $ are known as the cosmographic parameters. Using the obtained best-fit value of $ q $, we find that the present value of  $ j= -0.085238, ~-0.125, ~-0.097152 $ for the OHD, Pantheon and OHD+Pantheon data sets,  respectively.

\subsection{Statefinder Diagnostic}

In the literature, we find that to understand the universe's dynamics, geometric parameters play a vital role. When we study the deceleration parameter, we get information about the phase transition of the universe from deceleration to acceleration or vice versa. Therefore, it is required to study some additional higher-order derivatives of scale factor $ a $ like $ r $. To distinguish various dark energy models, we use a statefinder diagnostic technique defined in a pair $ \{r, s\} $, and is given in terms of $ q $ as \cite{Sahni:2002fz, Alam:2003sc}: 
\begin{equation} \label{29}
r=\frac{\dddot{a}}{H^{3}a}= 2 q^2 +q \text{, \ \ }s=\frac{-1+r}{3(-\frac{1}{2}+q)},
\end{equation}
where $ q\neq \frac{1}{2} $.

\begin{figure}\centering
	\subfloat[]{\label{sfa}\includegraphics[scale=0.54]{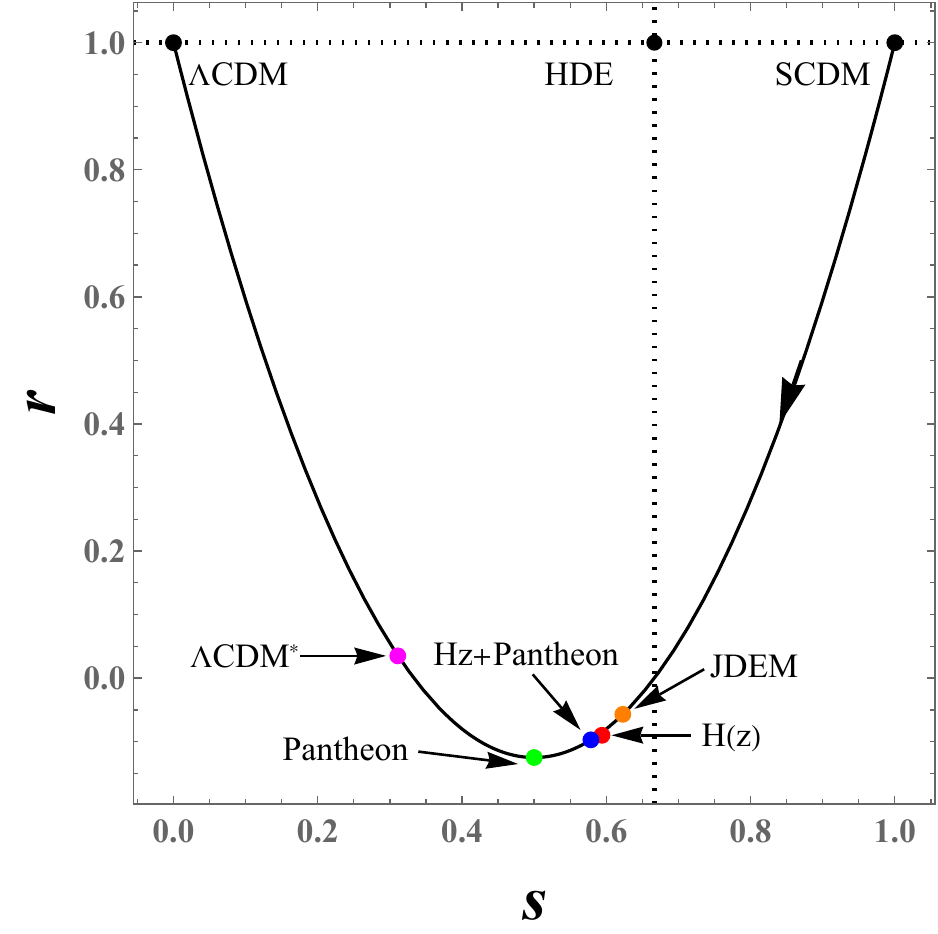}}\hfill
	\subfloat[]{\label{sfb}\includegraphics[scale=0.55]{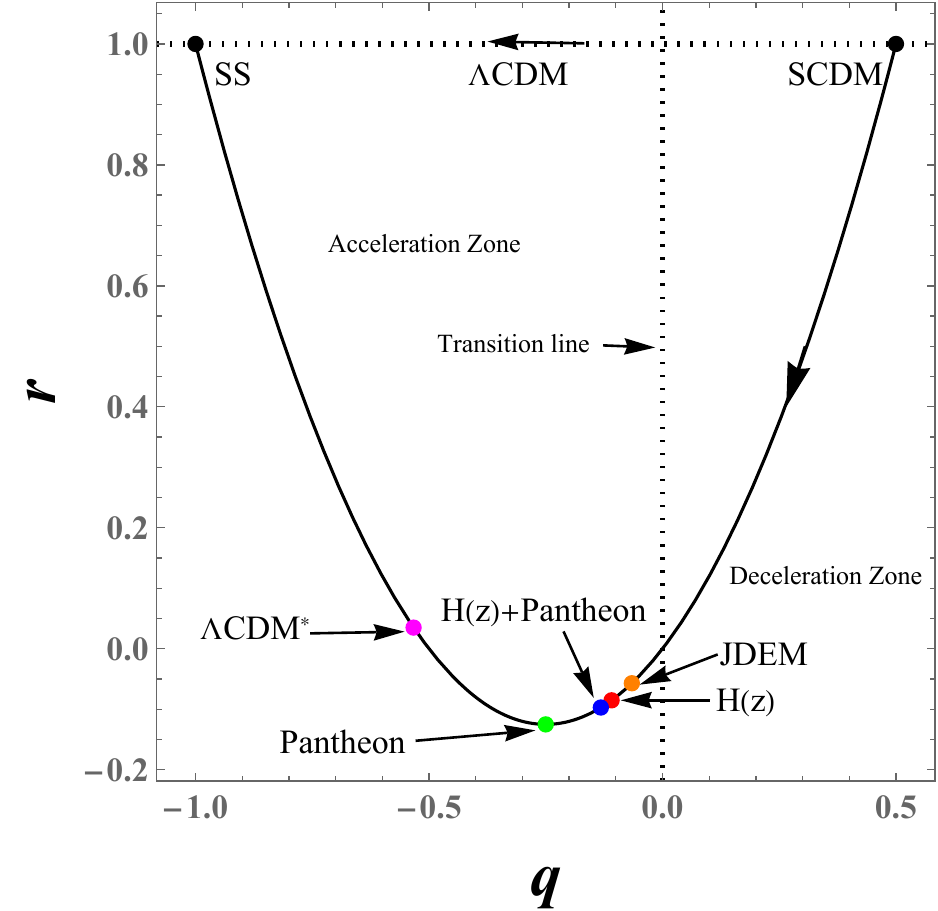}}	
\caption{ The $ s-r $ and $ q-r $ plots. }
\label{sf}
\end{figure}
Using this diagnostic approach, we observe the behavior of the model. We calculate the values of the parameters $ s $ and $ r $ parameters from the observed values of $ q $. Fig. \ref{sfa} shows the time evolution of the statefinder pair $\lbrace r,s \rbrace$ for the power-law cosmological model. The model converges to the fixed point ($ r=1, s=0 $) which corresponds to $ \Lambda $CDM. The point ($ r=1, s=2/3 $) corresponds to HDE. Fig. \ref{sfb} shows the time evolution of the statefinder pair $\lbrace r,q\rbrace$ for the said model. The point ($ r=1, q=0.5 $) corresponds to a matter-dominated Universe (SCDM) and converges to the point ($ r=1, q=-1 $) which corresponds to the de Sitter expansion (SS). The dots shown on the curves by arrows are the best-fit values of $ r $, $ s $, and $ q $ obtained by the latest OHD, Pantheon, and joint data (OHD+Pantheon). Here, our model approaches the $ \Lambda $CDM model as $ q \to -1 $ and is consistent with \cite{Rani:2014sia}.

\subsection{Om Diagnostic}

\qquad Here, we use the $ Om $ diagnostic technique known  to compare our model with the $ \Lambda $CDM model. This helps us distinguish various dark energy models without calculating the energy density and the EoS parameter. The pattern of the trajectories in the $ Om $ diagnostic plot indicates the various dark energy models. The definition of the  $ Om $ diagnostic in terms of  $ z $ is:  

\begin{equation}\label{40}
Om(z) = \frac{\lbrace\frac{H(z)}{H_0}\rbrace^2-1}{z^3+3z^2+3z}.
\end{equation}

\begin{figure}\centering
	\includegraphics[scale=0.45]{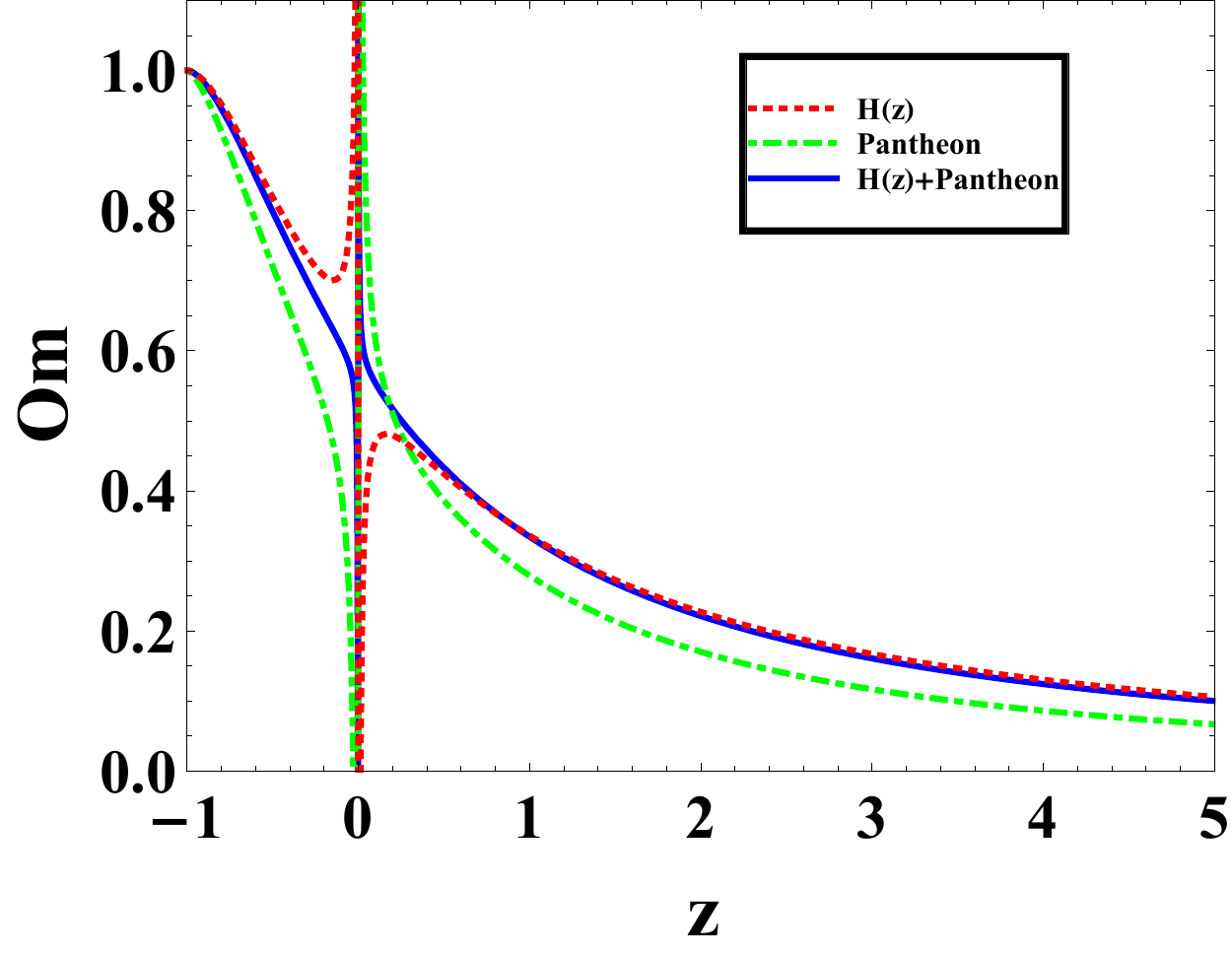}	
\caption{ The plot for om diagnostic.}
\label{om}
\end{figure}

The plots of the $ Om $ diagnostic help us to explain the nature of dark energy models. We know that if the curvature is positive  $ z $, then the model is a ghost dark energy model, if the curvature is negative concerning $ z $, we have quintessence, and if it has zero curvature, then the model represents the $ \Lambda $CDM model. Fig. \ref{om} shows that in late times, we have quintessence since the curvature is negative \textit{w.r.t.} $ z $ \cite{Singh:2019fpr, Singh:2022eun}.

\section{ Thermodynamical status of the model}
\qquad This section delves into applying the generalized second law of thermodynamics to our model. It is crucial to recognize that thermodynamics posits the entropy of isolated systems cannot decrease. We explore the calculation of the total entropy for this model, presuming that the apparent horizon is linked to temperature and entropy in a manner analogous to the black hole event horizon. According to the law, it mandates that the total entropy includes the entropy of all sources. As the Universe progresses, the rate of entropy change for both the contained fluid and the horizon must either stay constant or increase. To calculate the entropy on the boundary, we imagine a kind of spherical border around the universe, defined by its radius ($ R_h $). This radius is linked to the expansion rate of the universe ($ H $), which we have written in equation (\ref{14}) as \cite{Hawking:1975vcx, Pourbagher:2020zkm, Jamil:2009eb, Jamil:2010di, Saridakis:2020cqq}

\begin{equation}\label{18}
R_h = \frac{1}{H} = \frac{1}{H_0(1+z)^{(1+q)}}.
\end{equation}
Now, we calculate the amount of entropy at the edge of the horizon using the formula 
\begin{equation}\label{19}
S_{on}=\frac{\pi \kappa_b R_h^2}{l_{p}^2},
\end{equation}
where $ \kappa_b $ and $ l_{p} $ are the notations used for Boltzmann constant and Planck's length, respectively. Further, to investigate the matter entropy inside the horizon, the Gibbs relation is used as
\begin{equation}\label{20}
T_h dS_{in} = d(\rho V) + p dV = V d(\rho) + (\rho + p) dV,
\end{equation}
where $ p $ and $ \rho $ have been obtained from Eqs. (\ref{11}) and (\ref{12}) and the volume enclosed by the horizon is $ V=\frac{4\pi}{3} R_h^3 $. Here, $ T_h $ is the Hawking temperature on the edge of the horizon and can be calculated as
\begin{equation}\label{21}
T_h=\frac{1}{2\pi R_h}(1-\frac{\dot{R}_h}{2H R_h}).
\end{equation}
And, the condition for the second law of thermodynamics in thermodynamic equilibrium is as
\begin{equation}\label{22}
\dot{S}_{tot}= \dot{S}_{in} + \dot{S}_{on} \geq 0.
\end{equation}
 \begin{figure}\centering
	\subfloat[]{\label{sa}\includegraphics[scale=0.45]{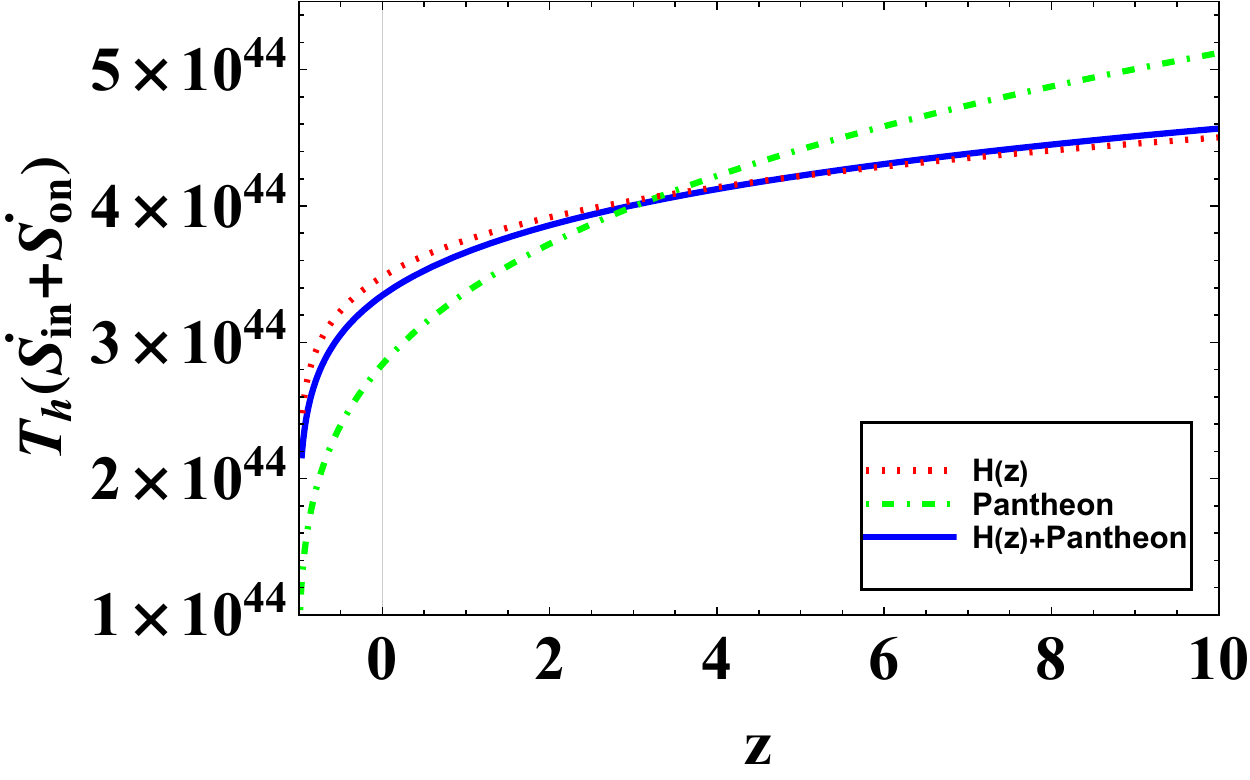}}\hfill
	\subfloat[]{\label{tb}\includegraphics[scale=0.39]{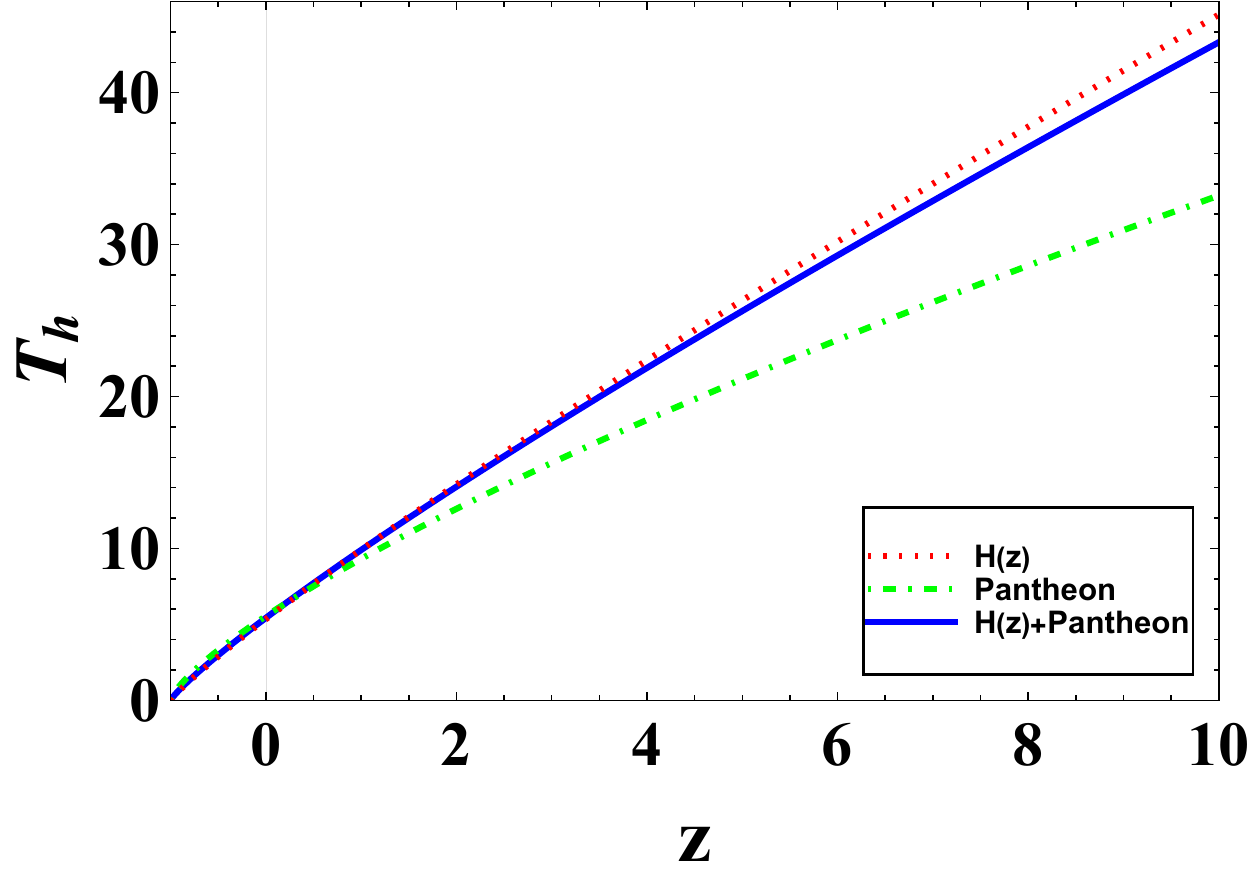}}
\caption{ The evolution of total entropy and temperature with respect to the redshift $ z $.}
\label{slot}
\end{figure}
To analyze the nature of total entropy and Hawkings temperature, a graphical representation has been applied by using the above relations and the value of $ H $, $ \rho $, and $ p $, we plot the trajectories for total entropy and Hawkings temperature (see Fig. \ref{slot}). Fig. \ref{sa} illustrates that The combined rates of the entropy change for matter and the horizon are always non-negative. Therefore, the Generalized Second Law of Thermodynamics is consistently satisfied for a spatially flat FLRW universe with pressureless dark matter within the apparent horizon. Also, Fig. \ref{tb} depicts that the Hawkings temperature decreases from early to late times.

\section{Conclusions}

The exactness of the cosmological observations extends the possibility of revealing the crucial properties of the Universe. In this work, the power-law cosmology stated in Eq. \ref{8} contains some outstanding features, making it different compared to the other models of the Universe. If we consider $ \zeta \geq1 $ in Eq. \ref{8}, then it directs the horizon, flatness, and age problems \cite{Mannheim:1989jh, Allen:1998vx, Kolb:1989bg}, and all these features are viable to the power-law cosmology to solve the cosmological constant problem analytically. In this work, we used the most recent observational data sets from OHD and Pantheon observations compiled in Table \ref{tabpan1}. We have obtained the constraints on two important cosmological parameters $ H_0 $ and $ q $ and compared our results with the earlier work \cite{Kumar:2011sw, Rani:2014sia}. Additionally, We have predicted these constraints with simulated data for large future surveys like JDEM. The numerical results have been concluded in Tables \ref{tab1}, and \ref{tabjdem}.

The late-time behaviour of our flat FLRW model in $ f(R,G) $ gravity has been studied, where $ F(R,G)=R+ \alpha e^{-G} $ and the invariant $ G $ is $ G=R^2-4R_{\mu\nu}R^{\mu\nu}+R_{\mu\nu\alpha\zeta}R^{\mu\nu\alpha\zeta} $. Since the field equations are difficult to solve in general, we assume a power law for the scale factor $ a $.   We constrain $ H $ and  $ q $ using recent observational datasets with the MCMC methodology and then proceed to study the behavior of the obtained model and the universe's evolution. The model has exhibited a singularity that is of the point type. The volume increases as $ t $ increases. The Hubble parameter monotonically decreases as $ z\rightarrow -1 $. The model is expanding with constant acceleration. The energy density of the model monotonically decreases with an increase in time, starting from infinity, and at late times, it tends to zero.  Fig. \ref{rho}, shows that our model is a quintessence dark energy model for all observational datasets.

The deceleration parameter is considered a free model parameter and its present values are constrained as 
$ q=-0.109_{-0.014}^{+0.014} $, $ q=-0.25_{-0.15}^{+0.15} $, and $ q=-0.132_{-0.014}^{+0.014} $ using  OHD, Pantheon, and   OHD+Pantheon data sets,  respectively, which deviate from the present day best-fit values (see Fig. \ref{hpc}). The NEC and DEC  energy conditions are satisfied, but the SEC does not hold for all the observational data sets, which support quintessence.  The stability of this model is verified by the parameters, as illustrated in  (see Fig. \ref{ec}). The late-time acceleration of the universe is supported by a violation of the SEC. The present value of the jerk cosmographic parameter is  $ j=-0.085238,~-0.125,~-0.097152 $ for the data sets OHD, Pantheon, and  OHD+Pantheon, respectively. This deviates from that of the  $ \Lambda $CDM. Our model has a quintessence type of behavior at late times. This is shown in the figure of $ Om(z) $ which is convex concerning the $ z $-axis and shows stability during the evolution of the universe up to late times except at present (see Fig. \ref{om}).

Now, by using the statefinder diagnostic approach, we investigated the behavior of the model and also checked the divergence and convergence of our model concerning the SCDM and $ \Lambda $CDM models. From the observed values of $ q $, we can calculate the values of $ s $ and $ r $ parameters. The best-fit values of $ r $, $ s $, and $ q $ are obtained by the latest OHD, Pantheon, and joint data (OHD+Pantheon) (see Fig. \ref{sf}). The model approaches the $ \Lambda $CDM model as $ q \to -1 $ and is consistent with \cite{Rani:2014sia}. Furthermore, the second law of thermodynamics is satisfied in this model. And, we find the decreasing behavior of temperature from early to late times (see Fig. \ref{slot}). Also, We notice that the Hubble parameter $ H_0 $ is best fitted whereas the deceleration parameter $ q $ does not fit well according to the recent observations. Thus, after reviewing the obtained results of our model, we see that our model starts with a Point-type singularity and behaves like an expanding accelerated dark energy model that is of the quintessence type and tends to the $ \Lambda $CDM model at late times.

According to the JDEM surveys, it can be concluded that the Universe is accelerating expanding but with smaller values of Hubble constant within the structure of power law cosmology. Finally, we can conclude that the power-law cosmology exhibits many outstanding features despite that it does not succeed in explaining the redshift-based transition from deceleration to acceleration of the Universe because the deceleration parameter $ q $ does not depend on the redshift and time. Thus, we can conclude that the power law cosmology does not fit well in dealing with all the cosmological challenges despite having several prominent features.

\vskip0.2in 
\section{Acknowledgment} 
The authors express their thanks to  Prof. Sushant G. Ghosh, CTP, Jamia Millia Islamia, New Delhi, India for fruitful discussions and suggestions. 
\vskip0.2in
\textbf{\noindent Data Availability Statement}
No Data is associated with the manuscript.


\begin{thebibliography}{99}

%\cite{Kumar:2011sw}
\bibitem{Kumar:2011sw}
S.~Kumar,
%``Observational constraints on Hubble constant and deceleration parameter in power-law cosmology,''
Mon. Not. Roy. Astron. Soc. \textbf{422} (2012), 2532-2538.
%doi:10.1111/j.1365-2966.2012.20810.x
%[arXiv:1109.6924 [gr-qc]].
%66 citations counted in INSPIRE as of 04 Nov 2022

\bibitem{Rani:2014sia}
S.~Rani, A.~Altaibayeva, M.~Shahalam, J.~K.~Singh and R.~Myrzakulov,
%``Constraints on cosmological parameters in power-law cosmology,''
JCAP \textbf{03} (2015), 031.
%doi:10.1088/1475-7516/2015/03/031
%[arXiv:1404.6522 [gr-qc]].
%45 citations counted in INSPIRE as of 04 Nov 2022

%\cite{Cognola:2007vq}
\bibitem{Cognola:2007vq}
G.~Cognola, M.~Gastaldi and S.~Zerbini,
%``On the stability of a class of modified gravitational models,''
Int. J. Theor. Phys. \textbf{47}, 898-910 (2008).
%doi:10.1007/s10773-007-9516-x
%[arXiv:gr-qc/0701138 [gr-qc]].
%127 citations counted in INSPIRE as of 03 Nov 2022

%\cite{Nojiri:2007bt}
\bibitem{Nojiri:2007bt}
S.~Nojiri, S.~D.~Odintsov and P.~V.~Tretyakov,
%``From inflation to dark energy in the non-minimal modified gravity,''
Prog. Theor. Phys. Suppl. \textbf{172} (2008), 81-89.
%doi:10.1143/PTPS.172.81
%[arXiv:0710.5232 [hep-th]].
%159 citations counted in INSPIRE as of 04 Nov 2022

%\cite{Bamba:2015jqa}
\bibitem{Bamba:2015jqa}
K.~Bamba,
%``Cosmological Issues in $F(T)$ Gravity Theory,''
[arXiv:1504.04457 [gr-qc]].
%8 citations counted in INSPIRE as of 03 Nov 2022

%\cite{Odintsov:2018nch}
\bibitem{Odintsov:2018nch}
S.~D.~Odintsov, V.~K.~Oikonomou and S.~Banerjee,
%``Dynamics of inflation and dark energy from $F(R,G)$ gravity,''
Nucl. Phys. B \textbf{938}, 935-956 (2019).
%doi:10.1016/j.nuclphysb.2018.07.013
%[arXiv:1807.00335 [gr-qc]].
%54 citations counted in INSPIRE as of 03 Nov 2022

%\cite{Alvarenga:2013syu}
\bibitem{Alvarenga:2013syu}
F.~G.~Alvarenga, A.~de la Cruz-Dombriz, M.~J.~S.~Houndjo, M.~E.~Rodrigues and D.~S\'aez-G\'omez,
%``Dynamics of scalar perturbations in $f(R,T)$ gravity,''
Phys. Rev. D \textbf{87}, no.10, 103526 (2013)
[erratum: Phys. Rev. D \textbf{87}, no.12, 129905 (2013)].
%doi:10.1103/PhysRevD.87.103526
%[arXiv:1302.1866 [gr-qc]].
%216 citations counted in INSPIRE as of 03 Nov 2022

%\cite{Sharif:2012zzd}
\bibitem{Sharif:2012zzd}
M.~Sharif and M.~Zubair,
%``Thermodynamics in f(R,T) Theory of Gravity,''
JCAP \textbf{03} (2012), 028.
%[erratum: JCAP \textbf{05} (2012), E01]
%doi:10.1088/1475-7516/2012/03/028
%[arXiv:1204.0848 [gr-qc]].
%233 citations counted in INSPIRE as of 04 Nov 2022

%\cite{Houndjo:2011tu}
\bibitem{Houndjo:2011tu}
M.~J.~S.~Houndjo,
%``Reconstruction of f(R, T) gravity describing matter dominated and accelerated phases,''
Int. J. Mod. Phys. D \textbf{21}, 1250003 (2012).
%doi:10.1142/S0218271812500034
%[arXiv:1107.3887 [astro-ph.CO]].
%275 citations counted in INSPIRE as of 03 Nov 2022

%\cite{Jamil:2011ptc}
\bibitem{Jamil:2011ptc}
M.~Jamil, D.~Momeni, M.~Raza and R.~Myrzakulov,
%``Reconstruction of some cosmological models in f(R,T) gravity,''
Eur. Phys. J. C \textbf{72}, 1999 (2012).
%doi:10.1140/epjc/s10052-012-1999-9
%[arXiv:1107.5807 [physics.gen-ph]].
%263 citations counted in INSPIRE as of 03 Nov 2022

%\cite{Yousaf:2016lls}
\bibitem{Yousaf:2016lls}
Z.~Yousaf, K.~Bamba and M.~Z.~u.~H.~Bhatti,
%``Causes of Irregular Energy Density in $f(R,T)$ Gravity,''
Phys. Rev. D \textbf{93}, no.12, 124048 (2016).
%doi:10.1103/PhysRevD.93.124048
%[arXiv:1606.00147 [gr-qc]].
%171 citations counted in INSPIRE as of 03 Nov 2022

%\cite{Godani:2018sbl}
\bibitem{Godani:2018sbl}
N.~Godani,
%``FRW cosmology in f(R,T) gravity,''
Int. J. Geom. Meth. Mod. Phys. \textbf{16}, no.02, 1950024 (2018).
%doi:10.1142/S0219887819500245
%20 citations counted in INSPIRE as of 03 Nov 2022

%\cite{Alves:2016iks}
\bibitem{Alves:2016iks}
M.~E.~S.~Alves, P.~H.~R.~S.~Moraes, J.~C.~N.~de Araujo and M.~Malheiro,
%``Gravitational waves in $f(R,T)$ and $f(R,T^\phi)$ theories of gravity,''
Phys. Rev. D \textbf{94}, no.2, 024032 (2016).
%doi:10.1103/PhysRevD.94.024032
%[arXiv:1604.03874 [gr-qc]].
%90 citations counted in INSPIRE as of 03 Nov 2022

%\cite{Sharma:2014zya}
\bibitem{Sharma:2014zya}
N.~K.~Sharma and J.~K.~Singh,
%``Bianchi Type-II String Cosmological Model with Magnetic Field in f (R, T) Gravity,''
Int. J. Theor. Phys. \textbf{53}, no.9, 2912-2922 (2014).
%doi:10.1007/s10773-014-2089-6
%25 citations counted in INSPIRE as of 03 Nov 2022

%\cite{Nagpal:2018uza}
\bibitem{Nagpal:2018uza}
R.~Nagpal, J.~K.~Singh and S.~Ayg\"un,
%``FLRW cosmological models with quark and strange quark matters in $f(R,T)$ gravity,''
Astrophys. Space Sci. \textbf{363}, 114 (2018).
%doi:10.1007/s10509-018-3335-9
%[arXiv:1805.06766 [physics.gen-ph]].
%20 citations counted in INSPIRE as of 03 Nov 2022

%\cite{Yousaf:2017hsq}
\bibitem{Yousaf:2017hsq}
Z.~Yousaf, M.~Ilyas and M.~Z.~Bhatti,
%``Influence of modification of gravity on spherical wormhole models,''
Mod. Phys. Lett. A \textbf{32}, no.30, 1750163 (2017).
%doi:10.1142/S0217732317501632
%36 citations counted in INSPIRE as of 03 Nov 2022

%\cite{Das:2016mxq}
\bibitem{Das:2016mxq}
A.~Das, F.~Rahaman, B.~K.~Guha and S.~Ray,
%``Compact stars in $f(R,\mathcal {T})$ gravity,''
Eur. Phys. J. C \textbf{76}, no.12, 654 (2016).
%doi:10.1140/epjc/s10052-016-4503-0
%[arXiv:1608.00566 [gr-qc]].
%98 citations counted in INSPIRE as of 03 Nov 2022

%\cite{Singh:2014kca}
\bibitem{Singh:2014kca}
J.~K.~Singh and N.~K.~Sharma,
%``Bianchi Type-II Dark Energy Model in f(R,T) Gravity,''
Int. J. Theor. Phys. \textbf{53}, 1424-1433 (2014).
%doi:10.1007/s10773-013-1939-y
%25 citations counted in INSPIRE as of 03 Nov 2022

%\cite{Nagpal:2018mpv}
\bibitem{Nagpal:2018mpv}
R.~Nagpal, S.~K.~J.~Pacif, J.~K.~Singh, K.~Bamba and A.~Beesham,
%``Analysis with observational constraints in $ \Lambda $-cosmology in $f(R, T)$ gravity,''
Eur. Phys. J. C \textbf{78}, no.11, 946 (2018).
%doi:10.1140/epjc/s10052-018-6403-y
%[arXiv:1805.03015 [physics.gen-ph]].
%25 citations counted in INSPIRE as of 03 Nov 2022


%\cite{Li:2007jm, Nojiri:2005jg, Nojiri:2005am, Cognola:2006eg, Elizalde:2010jx, Izumi:2014loa, Oikonomou:2016rrv, Kleidis:2017ftt, Shaily:2024rjq, Oikonomou:2015qha, Escofet:2015gpa, Makarenko:2017vuk, Bamba:2014mya, Makarenko:2016jsy, MontelongoGarcia:2010ip}

%\cite{Li:2007jm}
\bibitem{Li:2007jm}
B.~Li, J.~D.~Barrow and D.~F.~Mota,
%``The Cosmology of Modified Gauss-Bonnet Gravity,''
Phys. Rev. D \textbf{76}, 044027 (2007).
%doi:10.1103/PhysRevD.76.044027
%[arXiv:0705.3795 [gr-qc]].
%259 citations counted in INSPIRE as of 03 Nov 2022

%\cite{Nojiri:2005jg}
\bibitem{Nojiri:2005jg}
S.~Nojiri and S.~D.~Odintsov,
%``Modified Gauss-Bonnet theory as gravitational alternative for dark energy,''
Phys. Lett. B \textbf{631}, 1-6 (2005).
%doi:10.1016/j.physletb.2005.10.010
%[arXiv:hep-th/0508049 [hep-th]].
%871 citations counted in INSPIRE as of 03 Nov 2022

%\cite{Nojiri:2005am}
\bibitem{Nojiri:2005am}
S.~Nojiri, S.~D.~Odintsov and O.~G.~Gorbunova,
%``Dark energy problem: From phantom theory to modified Gauss-Bonnet gravity,''
J. Phys. A \textbf{39}, 6627-6634 (2006).
%doi:10.1088/0305-4470/39/21/S62
%[arXiv:hep-th/0510183 [hep-th]].
%186 citations counted in INSPIRE as of 03 Nov 2022

%\cite{Cognola:2006eg}
\bibitem{Cognola:2006eg}
G.~Cognola, E.~Elizalde, S.~Nojiri, S.~D.~Odintsov and S.~Zerbini,
%``Dark energy in modified Gauss-Bonnet gravity: Late-time acceleration and the hierarchy problem,''
Phys. Rev. D \textbf{73}, 084007 (2006).
%doi:10.1103/PhysRevD.73.084007
%[arXiv:hep-th/0601008 [hep-th]].
%633 citations counted in INSPIRE as of 03 Nov 2022



%\cite{Elizalde:2010jx}
\bibitem{Elizalde:2010jx}
E.~Elizalde, R.~Myrzakulov, V.~V.~Obukhov and D.~Saez-Gomez,
%``LambdaCDM epoch reconstruction from F(R,G) and modified Gauss-Bonnet gravities,''
Class. Quant. Grav. \textbf{27}, 095007 (2010).
%doi:10.1088/0264-9381/27/9/095007
%[arXiv:1001.3636 [gr-qc]].
%183 citations counted in INSPIRE as of 03 Nov 2022

%\cite{Izumi:2014loa}
\bibitem{Izumi:2014loa}
K.~Izumi,
%``Causal Structures in Gauss-Bonnet gravity,''
Phys. Rev. D \textbf{90}, no.4, 044037 (2014).
%doi:10.1103/PhysRevD.90.044037
%[arXiv:1406.0677 [gr-qc]].
%56 citations counted in INSPIRE as of 03 Nov 2022

%\cite{Oikonomou:2016rrv}
\bibitem{Oikonomou:2016rrv}
V.~K.~Oikonomou,
%``Gauss-Bonnet Cosmology Unifying Late and Early-time Acceleration Eras with Intermediate Eras,''
Astrophys. Space Sci. \textbf{361}, no.7, 211 (2016).
%doi:10.1007/s10509-016-2800-6
%[arXiv:1606.02164 [gr-qc]].
%16 citations counted in INSPIRE as of 03 Nov 2022

%\cite{Kleidis:2017ftt}
\bibitem{Kleidis:2017ftt}
K.~Kleidis and V.~K.~Oikonomou,
%``Loop quantum cosmology-corrected Gauss\textendash{}Bonnet singular cosmology,''
Int. J. Geom. Meth. Mod. Phys. \textbf{15}, no.04, 1850064 (2017).
%doi:10.1142/S0219887818500640
%[arXiv:1711.09270 [gr-qc]].
%15 citations counted in INSPIRE as of 03 Nov 2022

%\cite{Shaily:2024rjq}
\bibitem{Shaily:2024rjq}
Shaily, J.~K.~Singh and A.~Singh,
%``Bouncing Cosmology in f(R,G) Gravity with Thermodynamic Analysis,''
Fortsch. Phys. \textbf{72} (2024) no.6, 2300244.
%doi:10.1002/prop.202300244
%0 citations counted in INSPIRE as of 28 Jun 2024


%\cite{Oikonomou:2015qha}
\bibitem{Oikonomou:2015qha}
V.~K.~Oikonomou,
%``Singular Bouncing Cosmology from Gauss-Bonnet Modified Gravity,''
Phys. Rev. D \textbf{92}, no.12, 124027 (2015).
%doi:10.1103/PhysRevD.92.124027
%[arXiv:1509.05827 [gr-qc]].
%119 citations counted in INSPIRE as of 03 Nov 2022

%\cite{Escofet:2015gpa}
\bibitem{Escofet:2015gpa}
A.~Escofet and E.~Elizalde,
%``Gauss\textendash{}Bonnet modified gravity models with bouncing behavior,''
Mod. Phys. Lett. A \textbf{31}, no.17, 1650108 (2016).
%doi:10.1142/S021773231650108X
%[arXiv:1510.05848 [gr-qc]].
%29 citations counted in INSPIRE as of 03 Nov 2022

%\cite{Makarenko:2017vuk}
\bibitem{Makarenko:2017vuk}
A.~N.~Makarenko and A.~N.~Myagky,
%``The asymptotic behavior of bouncing cosmological models in $F(\mathcal{G})$ gravity theory,''
Int. J. Geom. Meth. Mod. Phys. \textbf{14}, no.10, 1750148 (2017).
%doi:10.1142/S0219887817501481
%[arXiv:1708.03592 [gr-qc]].
%11 citations counted in INSPIRE as of 03 Nov 2022

%\cite{Bamba:2014mya}
\bibitem{Bamba:2014mya}
K.~Bamba, A.~N.~Makarenko, A.~N.~Myagky and S.~D.~Odintsov,
%``Bouncing cosmology in modified Gauss-Bonnet gravity,''
Phys. Lett. B \textbf{732} (2014), 349-355.
%doi:10.1016/j.physletb.2014.04.004
%[arXiv:1403.3242 [hep-th]].
%107 citations counted in INSPIRE as of 04 Nov 2022

%\cite{Makarenko:2016jsy}
\bibitem{Makarenko:2016jsy}
A.~N.~Makarenko,
%``The role of Lagrange multiplier in Gauss\textendash{}Bonnet dark energy,''
Int. J. Geom. Meth. Mod. Phys. \textbf{13}, no.05, 1630006 (2016).
%doi:10.1142/S0219887816300063
%18 citations counted in INSPIRE as of 03 Nov 2022

%\cite{MontelongoGarcia:2010ip}
\bibitem{MontelongoGarcia:2010ip}
N.~Montelongo Garcia, F.~S.~N.~Lobo, J.~P.~Mimoso and T.~Harko,
%``f(G) modified gravity and the energy conditions,''
J. Phys. Conf. Ser. \textbf{314} (2011), 012056.
%doi:10.1088/1742-6596/314/1/012056
%[arXiv:1012.0953 [gr-qc]].
%27 citations counted in INSPIRE as of 04 Nov 2022

%\cite{Lohiya:1998tg}
\bibitem{Lohiya:1998tg}
D.~Lohiya and M.~Sethi,
%``A Program for a problem free cosmology within a framework of a rich class of scalar tensor theories,''
Class. Quant. Grav. \textbf{16}, 1545-1563 (1999).
%doi:10.1088/0264-9381/16/5/306
%[arXiv:gr-qc/9803054 [gr-qc]].
%29 citations counted in INSPIRE as of 03 Nov 2022

%\cite{Sethi:1999sq}
\bibitem{Sethi:1999sq}
M.~Sethi, A.~Batra and D.~Lohiya,
%``On 'observational constraints on power - law cosmologies',''
Phys. Rev. D \textbf{60}, 108301 (1999).
%doi:10.1103/PhysRevD.60.108301
%[arXiv:astro-ph/9903084 [astro-ph]].
%55 citations counted in INSPIRE as of 03 Nov 2022

%\cite{Batra:2000kur}
\bibitem{Batra:2000kur}
A.~Batra, D.~Lohiya, S.~Mahajan and A.~Mukherjee,
%``NUCLEOSYNTHESIS IN A UNIVERSE WITH A LINEARLY EVOLVING SCALE FACTOR,''
Int. J. Mod. Phys. D \textbf{9}, no.06, 757-773 (2000).
%doi:10.1142/S0218271800000682
%14 citations counted in INSPIRE as of 03 Nov 2022

%\cite{Gehlaut:2002mj}
\bibitem{Gehlaut:2002mj}
S.~Gehlaut, A.~Mukherjee, S.~Mahajan and D.~Lohiya,
%``A 'Freely coasting' universe,''
Spacetime and Substance \textbf{14} (2002), 152.
%[arXiv:astro-ph/0209209 [astro-ph]].
%10 citations counted in INSPIRE as of 04 Nov 2022

%\cite{Gehlaut:2003xi}
\bibitem{Gehlaut:2003xi}
S.~Gehlaut, P.~K.~Geetanjali and D.~Lohiya,
%``A Concordant ''freely coasting'' cosmology,''
[arXiv:astro-ph/0306448 [astro-ph]].
%20 citations counted in INSPIRE as of 04 Nov 2022

%\cite{Dev:2008ey}
\bibitem{Dev:2008ey}
A.~Dev, D.~Jain and D.~Lohiya,
%``Power law cosmology - a viable alternative,''
[arXiv:0804.3491 [astro-ph]].
%30 citations counted in INSPIRE as of 03 Nov 2022

%\cite{Dev:2002sz}
\bibitem{Dev:2002sz}
A.~Dev, M.~Safonova, D.~Jain and D.~Lohiya,
%``Cosmological tests for a linear coasting cosmology,''
Phys. Lett. B \textbf{548}, 12-18 (2002).
%doi:10.1016/S0370-2693(02)02814-9
%[arXiv:astro-ph/0204150 [astro-ph]].
%57 citations counted in INSPIRE as of 03 Nov 2022

%\cite{Zhu:2007tm}
\bibitem{Zhu:2007tm}
Z.~H.~Zhu, M.~Hu, J.~S.~Alcaniz and Y.~X.~Liu,
%``Testing power-law cosmology with galaxy clusters,''
Astron. Astrophys. \textbf{483} (2008), 15.
%doi:10.1051/0004-6361:20077797
%[arXiv:0712.3602 [astro-ph]].
%49 citations counted in INSPIRE as of 04 Nov 2022

%\cite{Sethi:2005au}
\bibitem{Sethi:2005au}
G.~Sethi, A.~Dev and D.~Jain,
%``Cosmological constraints on a power law Universe,''
Phys. Lett. B \textbf{624}, 135-140 (2005).
%doi:10.1016/j.physletb.2005.08.005
%[arXiv:astro-ph/0506255 [astro-ph]].
%60 citations counted in INSPIRE as of 03 Nov 2022

%\cite{Shafer:2015kda}
\bibitem{Shafer:2015kda}
D.~L.~Shafer,
%``Robust model comparison disfavors power law cosmology,''
Phys. Rev. D \textbf{91}, no.10, 103516 (2015).
%doi:10.1103/PhysRevD.91.103516
%[arXiv:1502.05416 [astro-ph.CO]].
%44 citations counted in INSPIRE as of 06 Mar 2023

%\cite{deHaro:2023lbq, Singh:2018xjv, Singh:2022jue, Aviles:2014rma, delaCruz-Dombriz:2016bqh, Capozziello:2019cav, Shaily:2024nmy, Balhara:2023mgj, Singh:2023gvf, Singh:2024ckh, Shaily:2024xho, Goswami:2023knh, Shaily:2022enj, Pawar:2024juv, Shabani:2016dhj, Singh:2022nfm}







%\cite{deHaro:2023lbq}
\bibitem{deHaro:2023lbq}
J.~de Haro, S.~Nojiri, S.~D.~Odintsov, V.~K.~Oikonomou and S.~Pan,
%``Finite-time cosmological singularities and the possible fate of the Universe,''
Phys. Rept. \textbf{1034}, 1-114 (2023).
%doi:10.1016/j.physrep.2023.09.003
%[arXiv:2309.07465 [gr-qc]].
%17 citations counted in INSPIRE as of 18 Feb 2024

%\cite{Singh:2018xjv}
\bibitem{Singh:2018xjv}
J.~K.~Singh, K.~Bamba, R.~Nagpal and S.~K.~J.~Pacif,
%``Bouncing cosmology in $f(R,T)$ gravity,''
Phys. Rev. D \textbf{97}, no.12, 123536 (2018).
%doi:10.1103/PhysRevD.97.123536
%[arXiv:1807.01157 [gr-qc]].
%73 citations counted in INSPIRE as of 18 Feb 2024

%\cite{Singh:2022jue}
\bibitem{Singh:2022jue}
J.~K.~Singh, H.~Balhara, K.~Bamba and J.~Jena,
%``Bouncing cosmology in modified gravity with higher-order curvature terms,''
JHEP \textbf{03}, 191 (2023)
[erratum: JHEP \textbf{04}, 049 (2023)].
%doi:10.1007/JHEP03(2023)191
%[arXiv:2206.12423 [gr-qc]].
%15 citations counted in INSPIRE as of 18 Feb 2024


%\cite{Aviles:2014rma}
\bibitem{Aviles:2014rma}
A.~Aviles, A.~Bravetti, S.~Capozziello and O.~Luongo,
%``Precision cosmology with Pad\'e rational approximations: Theoretical predictions versus observational limits,''
Phys. Rev. D \textbf{90}, no.4, 043531 (2014).
%doi:10.1103/PhysRevD.90.043531
%[arXiv:1405.6935 [gr-qc]].
%107 citations counted in INSPIRE as of 18 Feb 2024

%\cite{delaCruz-Dombriz:2016bqh}
\bibitem{delaCruz-Dombriz:2016bqh}
\'A.~de la Cruz-Dombriz, P.~K.~S.~Dunsby, O.~Luongo and L.~Reverberi,
%``Model-independent limits and constraints on extended theories of gravity from cosmic reconstruction techniques,''
JCAP \textbf{12}, 042 (2016).
%doi:10.1088/1475-7516/2016/12/042
%[arXiv:1608.03746 [gr-qc]].
%51 citations counted in INSPIRE as of 18 Feb 2024

%\cite{Capozziello:2019cav}
\bibitem{Capozziello:2019cav}
S.~Capozziello, R.~D'Agostino and O.~Luongo,
%``Extended Gravity Cosmography,''
Int. J. Mod. Phys. D \textbf{28}, no.10, 1930016 (2019).
%doi:10.1142/S0218271819300167
%[arXiv:1904.01427 [gr-qc]].
%295 citations counted in INSPIRE as of 18 Feb 2024

%\cite{Shaily:2024nmy, Balhara:2023mgj, Balhara:2023owb}

%\cite{Shaily:2024nmy}
\bibitem{Shaily:2024nmy}
Shaily, A.~Singh, J.~K.~Singh and S.~Ray,
%``Late time phantom characteristic of the model in $f(R,T)$ gravity with quadratic curvature term,''
[arXiv:2402.01780 [gr-qc]].
%1 citations counted in INSPIRE as of 18 Feb 2024

%\cite{Balhara:2023mgj}
\bibitem{Balhara:2023mgj}
H.~Balhara, J.~K.~Singh and E.~N.~Saridakis,
%``Observational constraints and cosmographic analysis of $f({T},{T}_{{G}})$ gravity and cosmology,''
[arXiv:2312.17277 [gr-qc]].
%1 citations counted in INSPIRE as of 18 Feb 2024

%\cite{Singh:2024ckh}
\bibitem{Singh:2024ckh}
J.~K.~Singh, Shaily, H.~Balhara, S.~G.~Ghosh and S.~D.~Maharaj,
%``EDSFD parameterization in f(R,T) gravity with linear curvature terms,''
Phys. Dark Univ. \textbf{45} (2024), 101513.
%doi:10.1016/j.dark.2024.101513
%0 citations counted in INSPIRE as of 16 May 2024

%\cite{Shaily:2024xho}
\bibitem{Shaily:2024xho}
Shaily, A.~Singh, J.~K.~Singh, S.~Hussain and R.~Myrzakulov,
%``Stability analysis of a dark energy model in Rastall gravity,''
[arXiv:2402.08709 [gr-qc]].
%2 citations counted in INSPIRE as of 24 Jun 2024

%\cite{Goswami:2023knh, Shaily:2022enj, Pawar:2024juv}
\bibitem{Goswami:2023knh}
G.~K.~Goswami, R.~Rani, J.~K.~Singh and A.~Pradhan,
%``FLRW cosmology in Weyl type f(Q) gravity and observational constraints,''
[arXiv:2309.01233 [gr-qc]].
%2 citations counted in INSPIRE as of 09 Jul 2024

%\cite{Shaily:2022enj}
\bibitem{Shaily:2022enj}
Shaily, J.~K.~Singh, J.~R.~L.~Santos and M.~Zeyauddin,
%``Dynamical analysis of a hyperbolic solution in scale-covariant theory,''
Int. J. Mod. Phys. D \textbf{33} (2024) no.05n06, 2450024.
%doi:10.1142/S021827182450024X
%[arXiv:2207.05076 [gr-qc]].
%11 citations counted in INSPIRE as of 09 Jul 2024

%\cite{Pawar:2024juv}
\bibitem{Pawar:2024juv}
D.~D.~Pawar, D.~K.~Raut, A.~P.~Nirwal, Shaily and J.~K.~Singh,
%``Observational constraints on the wet dark fluid model in the fractal gravity,''
Astron. Comput. \textbf{48} (2024), 100848.
%doi:10.1016/j.ascom.2024.100848
%0 citations counted in INSPIRE as of 09 Jul 2024

%\cite{Shabani:2016dhj}
\bibitem{Shabani:2016dhj}
H.~Shabani and A.~H.~Ziaie,
%``Stability of the Einstein static universe in $f(R,T)$ gravity,''
Eur. Phys. J. C \textbf{77} (2017) no.1, 31.
%doi:10.1140/epjc/s10052-017-4597-z
%[arXiv:1606.07959 [gr-qc]].
%77 citations counted in INSPIRE as of 26 May 2022

%\cite{Singh:2022nfm}
\bibitem{Singh:2022nfm}
J.~K.~Singh, Shaily, S.~Ram, J.~R.~L.~Santos and J.~A.~S.~Fortunato,
%``The constrained cosmological model in Lyra geometry,''
Int. J. Mod. Phys. D \textbf{32}, no.07, 2350040 (2023).
%doi:10.1142/S0218271823500402.
%[arXiv:2209.06859 [gr-qc]].
%6 citations counted in INSPIRE as of 27 Nov 2023


%\cite{Holsclaw:2010sk, SNAP:2004hke, lam:2003sc}
\bibitem{Holsclaw:2010sk}
T.~Holsclaw, U.~Alam, B.~Sanso, H.~Lee, K.~Heitmann, S.~Habib and D.~Higdon,
%``Nonparametric Dark Energy Reconstruction from Supernova Data,''
Phys. Rev. Lett. \textbf{105} (2010), 241302.
%doi:10.1103/PhysRevLett.105.241302
%[arXiv:1011.3079 [astro-ph.CO]].
%181 citations counted in INSPIRE as of 18 Jul 2024

%\cite{SNAP:2004hke}
\bibitem{SNAP:2004hke}
G.~Aldering \textit{et al.} [SNAP],
%``Supernova / Acceleration Probe: A Satellite Experiment to Study the Nature of the Dark Energy,''
[arXiv:astro-ph/0405232 [astro-ph]].
%160 citations counted in INSPIRE as of 18 Jul 2024

%\cite{Asvesta:2022fts}
\bibitem{Asvesta:2022fts}
K.~Asvesta, L.~Kazantzidis, L.~Perivolaropoulos and C.~G.~Tsagas,
%``Observational constraints on the deceleration parameter in a tilted universe,''
Mon. Not. Roy. Astron. Soc. \textbf{513}, no.2, 2394-2406 (2022).
%doi:10.1093/mnras/stac922
%[arXiv:2202.00962 [astro-ph.CO]].
%19 citations counted in INSPIRE as of 28 May 2024


%\cite{Riess:1998dv}
\bibitem{Riess:1998dv}
A.~G.~Riess, R.~P.~Kirshner, B.~P.~Schmidt, S.~Jha, P.~Challis, P.~M.~Garnavich, A.~A.~Esin, C.~Carpenter, R.~Grashius and R.~E.~Schild, \textit{et al.}
%``BV RI light curves for 22 type Ia supernovae,''
Astron. J. \textbf{117}, 707-724 (1999).
%doi:10.1086/300738
%[arXiv:astro-ph/9810291 [astro-ph]].
%642 citations counted in INSPIRE as of 28 May 2024

%\cite{Jha:2005jg}
\bibitem{Jha:2005jg}
S.~Jha, R.~P.~Kirshner, P.~Challis, P.~M.~Garnavich, T.~Matheson, A.~M.~Soderberg, G.~J.~M.~Graves, M.~Hicken, J.~F.~Alves and H.~G.~Arce, \textit{et al.}
%``Ubvri light curves of 44 type ia supernovae,''
Astron. J. \textbf{131}, 527-554 (2006).
%doi:10.1086/497989
%[arXiv:astro-ph/0509234 [astro-ph]].
%297 citations counted in INSPIRE as of 28 May 2024

%\cite{Hicken:2009df}
\bibitem{Hicken:2009df}
M.~Hicken, P.~Challis, S.~Jha, R.~P.~Kirsher, T.~Matheson, M.~Modjaz, A.~Rest and W.~M.~Wood-Vasey,
%``CfA3: 185 Type Ia Supernova Light Curves from the CfA,''
Astrophys. J. \textbf{700}, 331-357 (2009).
%doi:10.1088/0004-637X/700/1/331
%[arXiv:0901.4787 [astro-ph.CO]].
%412 citations counted in INSPIRE as of 28 May 2024

%\cite{Hicken:2009dk}
\bibitem{Hicken:2009dk}
M.~Hicken, W.~M.~Wood-Vasey, S.~Blondin, P.~Challis, S.~Jha, P.~L.~Kelly, A.~Rest and R.~P.~Kirshner,
%``Improved Dark Energy Constraints from \textasciitilde{}100 New CfA Supernova Type Ia Light Curves,''
Astrophys. J. \textbf{700}, 1097-1140 (2009).
%doi:10.1088/0004-637X/700/2/1097
%[arXiv:0901.4804 [astro-ph.CO]].
%794 citations counted in INSPIRE as of 28 May 2024

%\cite{Hicken:2012zr}
\bibitem{Hicken:2012zr}
M.~Hicken, P.~Challis, R.~P.~Kirshner, A.~Rest, C.~E.~Cramer, W.~M.~Wood-Vasey, G.~Bakos, P.~Berlind, W.~R.~Brown and N.~Caldwell, \textit{et al.}
%``CfA4: Light Curves for 94 Type Ia Supernovae,''
Astrophys. J. Suppl. \textbf{200}, 12 (2012).
%doi:10.1088/0067-0049/200/2/12
%[arXiv:1205.4493 [astro-ph.CO]].
%162 citations counted in INSPIRE as of 28 May 2024

%\cite{Contreras:2009nt}
\bibitem{Contreras:2009nt}
C.~Contreras, M.~Hamuy, M.~M.~Phillips, G.~Folatelli, N.~B.~Suntzeff, S.~E.~Persson, M.~Stritzinger, L.~Boldt, S.~Gonzalez and W.~Krzeminski, \textit{et al.}
%``The Carnegie Supernova Project: First Photometry Data Release of Low-Redshift Type Ia Supernovae,''
Astron. J. \textbf{139}, 519-539 (2010).
%doi:10.1088/0004-6256/139/2/519
%[arXiv:0910.3330 [astro-ph.CO]].
%259 citations counted in INSPIRE as of 28 May 2024

%\cite{SDSS:2014irn}
\bibitem{SDSS:2014irn}
M.~Sako \textit{et al.} [SDSS],
%``The Data Release of the Sloan Digital Sky Survey-II Supernova Survey,''
Publ. Astron. Soc. Pac. \textbf{130}, no.988, 064002 (2018).
%doi:10.1088/1538-3873/aab4e0
%[arXiv:1401.3317 [astro-ph.CO]].
%204 citations counted in INSPIRE as of 28 May 2024


%\cite{SNLS:2010pgl}
\bibitem{SNLS:2010pgl}
J.~Guy \textit{et al.} [SNLS],
%``The Supernova Legacy Survey 3-year sample: Type Ia Supernovae photometric distances and cosmological constraints,''
Astron. Astrophys. \textbf{523}, A7 (2010).
%doi:10.1051/0004-6361/201014468
%[arXiv:1010.4743 [astro-ph.CO]].
%424 citations counted in INSPIRE as of 28 May 2024

%\cite{Pan-STARRS1:2017jku}
\bibitem{Pan-STARRS1:2017jku}
D.~M.~Scolnic \textit{et al.} [Pan-STARRS1],
%``The Complete Light-curve Sample of Spectroscopically Confirmed SNe Ia from Pan-STARRS1 and Cosmological Constraints from the Combined Pantheon Sample,''
Astrophys. J. \textbf{859}, no.2, 101 (2018).
%doi:10.3847/1538-4357/aab9bb
%[arXiv:1710.00845 [astro-ph.CO]].
%2052 citations counted in INSPIRE as of 28 May 2024

%\cite{SupernovaCosmologyProject:2011ycw}
\bibitem{SupernovaCosmologyProject:2011ycw}
N.~Suzuki \textit{et al.} [Supernova Cosmology Project],
%``The Hubble Space Telescope Cluster Supernova Survey: V. Improving the Dark Energy Constraints Above z\ensuremath{>}1 and Building an Early-Type-Hosted Supernova Sample,''
Astrophys. J. \textbf{746}, 85 (2012).
%doi:10.1088/0004-637X/746/1/85
%[arXiv:1105.3470 [astro-ph.CO]].
%1556 citations counted in INSPIRE as of 02 Jun 2024

%\cite{Riess:2017lxs}
\bibitem{Riess:2017lxs}
A.~G.~Riess, S.~A.~Rodney, D.~M.~Scolnic, D.~L.~Shafer, L.~G.~Strolger, H.~C.~Ferguson, M.~Postman, O.~Graur, D.~Maoz and S.~W.~Jha, \textit{et al.}
%``Type Ia Supernova Distances at Redshift \ensuremath{>} 1.5 from the Hubble Space Telescope Multi-cycle Treasury Programs: The Early Expansion Rate,''
Astrophys. J. \textbf{853}, no.2, 126 (2018).
%doi:10.3847/1538-4357/aaa5a9
%[arXiv:1710.00844 [astro-ph.CO]].
%209 citations counted in INSPIRE as of 02 Jun 2024

%\cite{SupernovaSearchTeam:2004lze}
\bibitem{SupernovaSearchTeam:2004lze}
A.~G.~Riess \textit{et al.} [Supernova Search Team],
%``Type Ia supernova discoveries at z \ensuremath{>} 1 from the Hubble Space Telescope: Evidence for past deceleration and constraints on dark energy evolution,''
Astrophys. J. \textbf{607}, 665-687 (2004).
%doi:10.1086/383612
%[arXiv:astro-ph/0402512 [astro-ph]].
%3804 citations counted in INSPIRE as of 02 Jun 2024

%\cite{Riess:2006fw}
\bibitem{Riess:2006fw}
A.~G.~Riess, L.~G.~Strolger, S.~Casertano, H.~C.~Ferguson, B.~Mobasher, B.~Gold, P.~J.~Challis, A.~V.~Filippenko, S.~Jha and W.~Li, \textit{et al.}
%``New Hubble Space Telescope Discoveries of Type Ia Supernovae at z\ensuremath{>}=1: Narrowing Constraints on the Early Behavior of Dark Energy,''
Astrophys. J. \textbf{659}, 98-121 (2007).
%doi:10.1086/510378
%[arXiv:astro-ph/0611572 [astro-ph]].
%1697 citations counted in INSPIRE as of 02 Jun 2024

%\cite{Mamon:2016dlv}
\bibitem{Mamon:2016dlv}
A.~A.~Mamon and S.~Das,
%``A parametric reconstruction of the deceleration parameter,''
Eur. Phys. J. C \textbf{77} (2017) no.7, 495.
%doi:10.1140/epjc/s10052-017-5066-4
%[arXiv:1610.07337 [gr-qc]].
%83 citations counted in INSPIRE as of 10 Jul 2024

%\cite{Planck:2018vyg}
\bibitem{Planck:2018vyg}
N.~Aghanim \textit{et al.} [Planck],
%``Planck 2018 results. VI. Cosmological parameters,''
Astron. Astrophys. \textbf{641} (2020), A6
[erratum: Astron. Astrophys. \textbf{652} (2021), C4].
%doi:10.1051/0004-6361/201833910
%[arXiv:1807.06209 [astro-ph.CO]].
%14316 citations counted in INSPIRE as of 16 Jul 2024

%\cite{Alam:2003sc}
\bibitem{Alam:2003sc}
U.~Alam, V.~Sahni, T.~D.~Saini and A.~A.~Starobinsky,
%``Exploring the expanding universe and dark energy using the Statefinder diagnostic,''
Mon. Not. Roy. Astron. Soc. \textbf{344} (2003), 1057.
%doi:10.1046/j.1365-8711.2003.06871.x
%[arXiv:astro-ph/0303009 [astro-ph]].
%753 citations counted in INSPIRE as of 18 Jul 2024

%\cite{Gelman:1992zz, Brooks:1998}
\bibitem{Gelman:1992zz}
A.~Gelman and D.~B.~Rubin,
%``Inference from Iterative Simulation Using Multiple Sequences,''
Statist. Sci. \textbf{7} (1992), 457-472.
%doi:10.1214/ss/1177011136
%946 citations counted in INSPIRE as of 08 Mar 2023

\bibitem{Brooks:1998} Brooks, S. P., and A. Gelman. 1998. General methods for monitoring convergence of iterative simulations, Journal of Computational and Graphical Statistics 7: 434–455.

%\cite{Singh:2018cip}
\bibitem{Singh:2018cip}
C.~P.~Singh and M.~Srivastava,
%``Viscous cosmology in new holographic dark energy model and the cosmic acceleration,''
Eur. Phys. J. C \textbf{78}, no.3, 190 (2018)
%doi:10.1140/epjc/s10052-018-5683-6
%13 citations counted in INSPIRE as of 01 Mar 2023

%\cite{Sahu:2016ccd}
\bibitem{Sahu:2016ccd}
S.~K.~Sahu, S.~K.~Tripathy, P.~K.~Sahoo and A.~Nath,
%``Cosmic Transit and Anisotropic Models in f(R,T) Gravity,''
Chin. J. Phys. \textbf{55}, 862-869 (2017)
%doi:10.1016/j.cjph.2017.02.013
%[arXiv:1611.03476 [gr-qc]].
%28 citations counted in INSPIRE as of 01 Mar 2023

\bibitem{Curiel:2014zba}
E.~Curiel,
%``A Primer on Energy Conditions,''
Einstein Stud. \textbf{13}, 43-104 (2017).
%doi:10.1007/978-1-4939-3210-8\_3
%[arXiv:1405.0403 [physics.hist-ph]].

%\cite{Kontou:2020bta}
\bibitem{Kontou:2020bta}
E.~A.~Kontou and K.~Sanders,
%``Energy conditions in general relativity and quantum field theory,''
Class. Quant. Grav. \textbf{37} (2020) no.19, 193001.
%doi:10.1088/1361-6382/ab8fcf
%[arXiv:2003.01815 [gr-qc]].

\bibitem{Visser:1995cc}
M.~Visser, ``Lorentzian wormholes: From Einstein to Hawking'', 1995.

\bibitem{Caldwell:1999ew}
R.~R.~Caldwell,
%``A Phantom menace?,''
Phys. Lett. B \textbf{545}, 23-29 (2002).
%doi:10.1016/S0370-2693(02)02589-3
%[arXiv:astro-ph/9908168 [astro-ph]].

\bibitem{Rubakov:2014jja}
V.~A.~Rubakov,
%``The Null Energy Condition and its violation,''
Phys. Usp. \textbf{57} (2014), 128-142.
%doi:10.3367/UFNe.0184.201402b.0137
%[arXiv:1401.4024 [hep-th]].

\bibitem{Flanagan:1996gw}
E.~E.~Flanagan and R.~M.~Wald,
%``Does back reaction enforce the averaged null energy condition in semiclassical gravity?,''
Phys. Rev. D \textbf{54} (1996), 6233-6283.
%doi:10.1103/PhysRevD.54.6233
%[arXiv:gr-qc/9602052 [gr-qc]].

\bibitem{Penrose:1964wq}
R.~Penrose,
%``Gravitational collapse and space-time singularities,''
Phys. Rev. Lett. \textbf{14} (1965), 57-59.
%doi:10.1103/PhysRevLett.14.57

%\cite{Sahni:2002fz}
\bibitem{Sahni:2002fz}
V.~Sahni, T.~D.~Saini, A.~A.~Starobinsky and U.~Alam,
%``Statefinder: A New geometrical diagnostic of dark energy,''
JETP Lett. \textbf{77}, 201-206 (2003).
%doi:10.1134/1.1574831
%[arXiv:astro-ph/0201498 [astro-ph]].
%915 citations counted in INSPIRE as of 03 Nov 2022

%\cite{Singh:2019fpr}
\bibitem{Singh:2019fpr}
J.~K.~Singh and R.~Nagpal,
%``FLRW cosmology with EDSFD parametrization,''
Eur. Phys. J. C \textbf{80} (2020) no.4, 295.
%doi:10.1140/epjc/s10052-020-7827-8
%[arXiv:1910.09289 [physics.gen-ph]].
%8 citations counted in INSPIRE as of 06 Apr 2023

%\cite{Singh:2022eun}
\bibitem{Singh:2022eun}
J.~K.~Singh, A.~Singh, G.~K.~Goswami and J.~Jena,
%``Dynamics of a parametrized dark energy model in f(R,T) gravity,''
Annals Phys. \textbf{443} (2022), 168958.
%doi:10.1016/j.aop.2022.168958
%[arXiv:2204.07599 [gr-qc]]
%0 citations counted in INSPIRE as of 23 Jun 2022

\bibitem{Hawking:1975vcx}
S.~W.~Hawking,
%``Particle Creation by Black Holes,''
Commun. Math. Phys. \textbf{43}, 199-220 (1975)
[erratum: Commun. Math. Phys. \textbf{46}, 206 (1976)].
%doi:10.1007/BF02345020
%9665 citations counted in INSPIRE as of 10 Jun 2022

%\cite{Pourbagher:2020zkm}
\bibitem{Pourbagher:2020zkm}
A.~Pourbagher and A.~Amani,
%``Thermodynamics of the viscous $f(T, B)$ gravity in the new agegraphic dark energy model,''
Mod. Phys. Lett. A \textbf{35}, no.20, 2050166 (2020).
%doi:10.1142/S0217732320501667
%[arXiv:2006.09172 [gr-qc]].
%10 citations counted in INSPIRE as of 10 Jun 2022

%\cite{Singh:2023ryd, Shrivastava:2021hsu, Singh:2022jue, Goswami:2022vfq, Singh:2022gln, Singh:2022ptu, Singh:2022eun, Singh:2019fpr, Singh:2022nfm, Shaily:2022enj, Bhardwaj:2022lrm, Singh:2023gxd, Singh:2022wwa}

%\cite{Jamil:2009eb}
\bibitem{Jamil:2009eb}
M.~Jamil, E.~N.~Saridakis and M.~R.~Setare,
%``Thermodynamics of dark energy interacting with dark matter and radiation,''
Phys. Rev. D \textbf{81}, 023007 (2010).
%doi:10.1103/PhysRevD.81.023007
%[arXiv:0910.0822 [hep-th]].
%186 citations counted in INSPIRE as of 06 Mar 2024

%\cite{Jamil:2010di}
\bibitem{Jamil:2010di}
M.~Jamil, E.~N.~Saridakis and M.~R.~Setare,
%``The generalized second law of thermodynamics in Horava-Lifshitz cosmology,''
JCAP \textbf{11}, 032 (2010).
%doi:10.1088/1475-7516/2010/11/032
%[arXiv:1003.0876 [hep-th]].
%122 citations counted in INSPIRE as of 06 Mar 2024

%\cite{Saridakis:2020cqq}
\bibitem{Saridakis:2020cqq}
E.~N.~Saridakis and S.~Basilakos,
%``The generalized second law of thermodynamics with Barrow entropy,''
Eur. Phys. J. C \textbf{81}, no.7, 644 (2021).
%doi:10.1140/epjc/s10052-021-09431-y
%[arXiv:2005.08258 [gr-qc]].
%76 citations counted in INSPIRE as of 06 Mar 2024

%\cite{Mannheim:1989jh, Allen:1998vx, Kolb:1989bg}
\bibitem{Mannheim:1989jh}
P.~D.~Mannheim,
%``Conformal Cosmology With No Cosmological Constant,''
Gen. Rel. Grav. \textbf{22} (1990), 289-298.
%doi:10.1007/BF00756278
%136 citations counted in INSPIRE as of 18 Jul 2024


%\cite{Allen:1998vx}
\bibitem{Allen:1998vx}
R.~E.~Allen,
%``Four testable predictions of instanton cosmology,''
AIP Conf. Proc. \textbf{478} (1999) no.1, 204-207.
%doi:10.1063/1.59392
%[arXiv:astro-ph/9902042 [astro-ph]].
%24 citations counted in INSPIRE as of 18 Jul 2024

%\cite{Kolb:1989bg}
\bibitem{Kolb:1989bg}
E.~W.~Kolb,
%``A Coasting Cosmology,''
Astrophys. J. \textbf{344} (1989), 543-550.
%doi:10.1086/167825
%120 citations counted in INSPIRE as of 18 Jul 2024

\end{thebibliography}
\end{document}